\documentclass{article}

\usepackage{arxiv}

\usepackage[utf8]{inputenc} 
\usepackage[T1]{fontenc}    
\usepackage{hyperref}       
\usepackage{url}            
\usepackage{booktabs}       
\usepackage{amsfonts}       
\usepackage{nicefrac}       
\usepackage{microtype}      
\usepackage{lipsum}		
\usepackage{graphicx}
\usepackage{natbib}
\usepackage{doi}
\usepackage{xcolor}
\usepackage{amsmath}
\usepackage{soul}
\usepackage{graphicx}
\usepackage{tikz}
\usepackage{booktabs}
\usepackage{amssymb}

\usepackage{gensymb}

\usepackage{adjustbox}
\usepackage{ulem}
\usepackage{cancel}
\usepackage{svg}
\usepackage{amsmath}

\usepackage{caption}
\usepackage{graphicx}
\usepackage{comment}
\DeclareCaptionLabelFormat{adja-page}{\hrulefill\\#1 #2 \emph{(previous page)}}

\title{Aqueous foams in microgravity, measuring bubble sizes}

\author{ 
\hspace{1mm}Marina Pasquet$^1$,
\hspace{1mm}Nicolo Galvani$^{2,3}$,
\hspace{1mm}Olivier Pitois$^3$,
\hspace{1mm}Sylvie Cohen-Addad$^{2,3}$,
\hspace{1mm}Reinhard Höhler$^{2,3}$,\\
\hspace{1mm}\textbf{Anthony T. Chieco$^4$},
\hspace{1mm}\textbf{Sam Dillavou$^4$},
\hspace{1mm}\textbf{Jesse M. Hanlan$^4$},
\hspace{1mm}\textbf{Douglas J. Durian$^4$},\\
\hspace{1mm}\textbf{Emmanuelle Rio$^1$},
\hspace{1mm}\textbf{Anniina Salonen$^1$},
\hspace{1mm}\textbf{Dominique Langevin$^1$},\\
	$^1$ Universit\'e Paris-Saclay, CNRS, Laboratoire de Physique des Solides, 91405, Orsay, France.\\
	$^2$ Sorbonne Université, Institut des NanoSciences de Paris, Paris, France.\\
    $^3$ Université Gustave Eiffel, Laboratoire Navier, Marne-la-Vallée, France. \\
    $^4$ Department of Physics and Astronomy, University of Pennsylvania, Philadelphia, Pennsylvania 19104, USA. 
}

\renewcommand{\shorttitle}{Aqueous foams in microgravity, measuring bubble sizes}

\hypersetup{
pdftitle={Aqueous foams in microgravity, measuring bubble sizes},
}

\newcommand{\Reinhard}[1]{\textcolor{black}{#1}}
\newcommand{\Sylvie}[1]{\textcolor{black}{#1}}
\newcommand{\Manue}[1]{\textcolor{black}{#1}}
\newcommand{\Dominique}[1]{\textcolor{black}{#1}}
\newcommand{\DJD}[1]{\textcolor{black}{#1}}
\newcommand{\Marina}[1]{\textcolor{black}{#1}}
\newcommand{\Anniina}[1]{\textcolor{black}{#1}}

\newcommand{\Anthony}[1]{\textcolor{black}{#1}}
\newcommand{\Nicolo}[1]{\textcolor{black}{#1}}

\begin{document}

\maketitle

\begin{abstract}
The paper describes a study of wet foams in microgravity whose bubble size distribution evolves due to diffusive gas exchange. We focus on the comparison between the size of bubbles determined from images of the foam surface and the size of bubbles in the bulk foam, determined from Diffuse Transmission Spectroscopy (DTS). Extracting the bubble size distribution from images of a foam surface is difficult so we have used three different procedures : manual analysis, automatic analysis with a customized Python script and machine learning analysis. Once various pitfalls were identified and taken into account, all the three procedures yield identical results within error bars. DTS only allows the determination of  an average bubble radius which is proportional to the photon transport mean free path $\ell^*$. The relation between the measured diffuse transmitted light intensity and {$\ell^*$}  previously derived for  slab-shaped samples of infinite lateral extent does not apply to the cuboid geometry of the cells used in the microgravity experiment. A new more general expression of the diffuse intensity transmitted  with specific optical boundary conditions has been derived and applied to determine the average bubble radius. The temporal evolution of the average bubble radii deduced from DTS and of the same average radii of the bubbles measured at the sample surface \Dominique{is the same (to a factor probably close to one)} throughout the coarsening. Finally, ground experiments were performed to compare bubble size distributions  in a bulk wet foam and at its surface at times so short that diffusive gas exchange is insignificant. They were found to be similar, confirming that bubbles seen at the surface are representative of the bulk foam bubbles.
\end{abstract}


\section{Introduction}
Foams are dispersions of gas in liquid or solid matrices~\cite{Cantat2013, Gibson1997}. They have applications in many different fields, liquid foams in detergency, food, medicine, fire-fighting, oil recovery, solid foams in aerated materials for packaging, thermal and phonic insulation in building constructions for instance. Solid foams are often obtained from liquid foams containing large amounts of liquid, hence a better knowledge of the properties of such ``wet" foams will help to optimize their manufacturing processes, which are to date still mostly empirical. Indeed, the matrix is initially liquid and later solidified and the liquid content of the wet foams drains very rapidly due to gravity, making it difficult to explore their properties experimentally. Hence knowledge of the properties of foams containing more than a few percent of liquid is extremely limited.
To study these wet foams, we designed\ experiments for microgravity conditions in the International Space Station (ISS). We have studied foams with liquid fractions between 15 and 50\%.
Above 36\%, the bubbles are disconnected and spherical and the dispersion is called a bubbly liquid. Below 36\%, the bubbles are pressed together and distort from spherical with thin liquid films formed between them.

The ISS project is focused on the study of wet foam coarsening due to gas transfer between bubbles with different internal Laplace pressures. In dry foams, gas transfer occurs mainly through the liquid films between bubbles;it has been shown that in this case, the bubble radius increases as the square root of time \cite{Mullins1986}. 
In bubbly liquids, gas is transferred through regions of size comparable to the bubble size, and the radius then increases as the cubic root of time. This is the well-known regime of Ostwald ripening~\cite{Taylor1998}.  \Anniina{The goal of the ISS experiments is to study the growth law of the average bubble size distribution during the coarsening, as a function of liquid fraction. However, given the constrictions of the measuring geometry this opens questions on the impact of bubble analysis in such confined geometries. }

The module designed for the foam study in the ISS combines different diagnostics: An overview camera to take
images of the foam  surface, as well as a laser and a detector to measure the  intensity of the diffuse light transmitted through the sample~\cite{Born}. The average radius of bubbles in the bulk  is determined from Diffuse Transmission Spectroscopy (DTS), while the bubble size distribution and average radii at the surface are obtained from planimetric measurements of the contour area of the bubbles touching the sample cell wall.

Several biases may be encountered in such planimetric measurements, leading to average sizes and distributions deduced from surface observations unrepresentative of the bulk structure. \Manue{For instance, numerical simulations were performed for disordered totally dry (0\% liquid fraction) 3D foams constrained against a wall and the  distribution of bubbles radius measured at the surface and in the bulk were found to be different~\cite{Wang2009}}. 
Besides this statistical effect, other possible experimental biases have been pointed out by Cheng and Lemlich~\cite{Cheng1983} such  as: i) bubble segregation, if small bubbles tend to accumulate at the wall and wedge large bubbles away from the surface; ii) bubble distortion, as the surface bubbles are squeezed against the wall due to the confinement pressure exerted by the underlying bubble layers; iii) local coarsening process, if the surface modifies the Laplace-driven gas diffusion exchanges between bubbles.
Therefore, for our coarsening study of wet foams in the present paper,  we study the extent to which surface bubble size distributions and average radii are representative of the foam bulk structure.

\par
We present the samples and the experimental setup used in the ISS and on ground in section 2; then we describe in section 3  three different image analysis procedures that we used for the planimetric measurements to determine the average bubble size and distribution at the surface. We present in section 4, a ground experiment performed to compare the initial bubble size distribution measured at the surface of a wet foam right after its production to that measured for a bubble monolayer of the same foam. In the case of such a layer, the bubble sizes can be determined without any ambiguity.

Then we turn to the analysis of the DTS data that yield an average of the bubble size.
Previous  expressions relating the average bubble size to the transmitted intensity published in the literature~\cite{durian1991}  could not be used, due to the unusual geometry of our ISS sample cells. 
We have therefore derived the required analytical relation starting from the light diffusion equation, and taking into account the cuboid shape of the cell and its specific optical boundary conditions. This calculation is described in section 5. The  analysis of the ISS experimental data using this model is performed all along the coarsening process and presented in section 6.

\section{Experimental}
 \label{sec:Experimental}
 
\subsection{ISS Sample composition}
\label{sec:ISSsample}
The foams were made \Sylvie{by mixing air} with aqueous solutions of a\Dominique{n} ionic surfactant, tetradecyl-trimethyl-ammonium bromide (TTAB \Sylvie{with concentration 5 g/l}). This surfactant does not evolve chemically in water, hence it was chosen because of the long storage periods imposed by experiments in the ISS. The water was ultrapure water from a Millipore device (resistivity 18 M$\Omega$).  
\Anniina{Ten different liquid fractions were studied:}
15.2, 20.3, 25.4, 30.6, 32.5, 35.1, 37.9, 40.2, 45.3 and 50.2\%. \Anniina{The weights and volumes of all the cells were measured and the foaming liquids were weighed using a precision balance during filling. This results in the high precision in the value of the liquid fraction.} In the text, \Anniina{the liquid fraction values are rounded to the nearest integer} to facilitate the reading.

\subsection{Experiments on the ISS}
\label{sec:ExperimentalISS}
\begin{figure}[tbp]
 \includegraphics[width=1\linewidth]{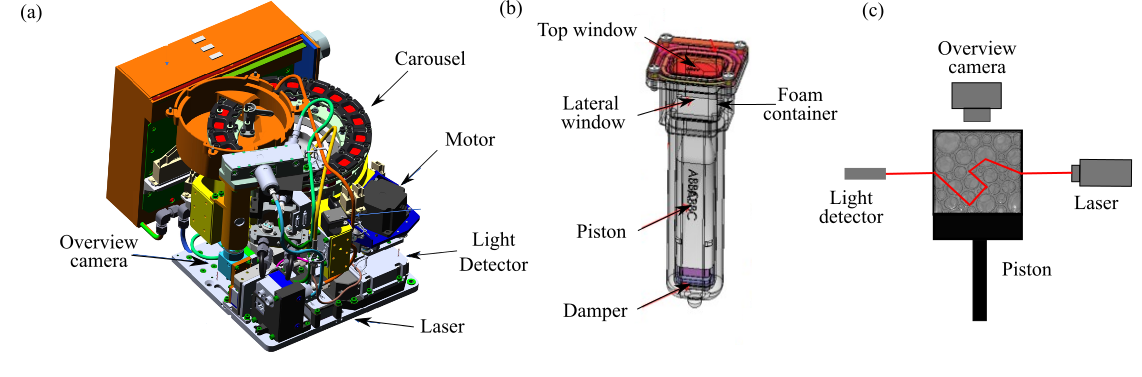}
 \caption{Figure 1. (a) ISS module showing the carousel of size 40 cm x 28 cm x 27 cm containing the 20 cells and the optical diagnostics. (b) A foam sample cell showing the location of the piston used to produce the foam in the lower part and the measurement region in the upper part. (c) Schema of the overview camera and the Diffuse Transmission Spectroscopy probe in the ISS instrument. }
 \label{fig:exmp}
 \end{figure}

The experimental setup in the ISS module is illustrated schematically in Figure \ref{fig:exmp}(a). More details can be found in reference~\cite{Born}.

 The module contains a carousel holding the foam sample cells, which were filled with different volumes of the solution in order to set the liquid fractions. The cells are made of COC (Cyclic Olefin Copolymer), a transparent plastic material which is impermeable to water. The lower part of each cell contains a piston which can be \st{moved} \Sylvie{translated up and and down} using magnetic forces to produce the foam.
 \Sylvie{The upper part of each cell, of cuboid shape (dimensions $\simeq$ 1 cm, Figure~\ref{fig:exmp}b) is the observation chamber, where microscopy and optical measurements are performed. Bubbles are formed as the liquid and the gas are forced by the piston motion to mix through the thin gap between the lateral sides of the piston and the walls of the chamber.}  The piston is actuated periodically \Sylvie{for a few minutes} by a magnetic field, producing foam with average bubble radius of about 50 $\mu$m. When the study of a sample cell is terminated, the carousel rotates to move a new cell in front of the laser and the various cameras. After the piston is stopped, the overview camera begins recording images of the foam at the top window of the cell. 
 A laser illuminates the foam at the center of the upper part of the cell through the lateral window (Figure~\ref{fig:exmp}b). Then light is multiply scattered by the gas-liquid interfaces.  Various measurements of the diffuse transmitted or backscattered light can be performed with the instrument~\cite{Born}. Here we focus on the Diffuse Transmission Spectroscopy (DTS) probe schematically described in Figure~\ref{fig:exmp}c).

\subsection{Ground based experiments}
\label{sec:Experimentalground}
To compare the bubble size distribution in the bulk and at the surface of a foam, we have developed a ground experiment. 
To study foams similar to those of the ISS experiments, we used a foaming liquid constituted of TTAB surfactant (5~g/L) dissolved in ultrapure water (MilliQ), and produced foams using the double syringe method \cite{Gaillard2017}. A  syringe is filled with the liquid, while a second identical syringe is filled with air saturated with perfluorohexane vapor, which is used here to strongly slow down coarsening \cite{Cohen-Addad2004}, in proportion to achieve the desired liquid fraction in the final foam. Then the two syringes are connected with a rigid connector (Combifix Adapter Luer-Luer) and put on a support which fixes the piston positions. To produce foam the jointed bodies of the syringes are then moved back and forth, at 1~Hz for 2~minutes. Less than 60~s after the end of the foaming process, a sample cell with transparent glass windows, (inner thickness 2~mm, much larger than the bubble size),  is filled with foam. Simultaneously, 1~$\mu$L of the same foam is injected in a second cell, previously filled with the foaming liquid. It also has transparent glass windows; the one at the top is horizontal.  The bubbles are dispersed in this liquid and form a dilute bubble monolayer at the upper window.  The surface of the foam sample and the bubble monolayer are observed using a video-microscope with illumination by transmission. The first images are taken within two minutes after the end of the foaming process. 

\section{Image analysis}
 \label{sec:img_analysis}
\begin{figure}[!ht]
 \includegraphics[width=1\linewidth]{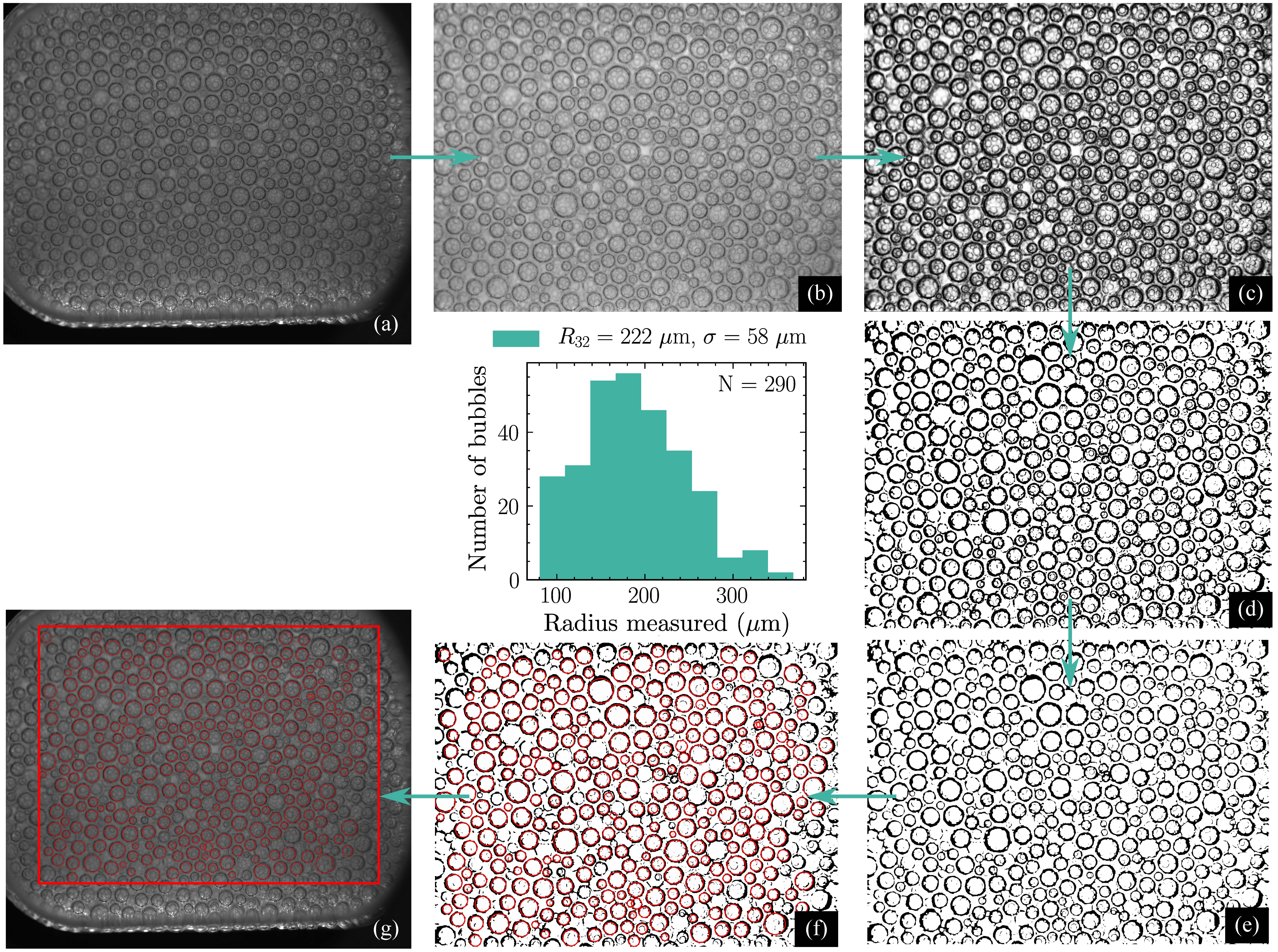}
 \caption{Foam with a liquid fraction of 40\%  (a) raw image, (b) image whose brightness and contrast have been increased, (c) image whose contrast has again been improved and this time locally, (d) binarized image using a local thresholding method, (e) image having undergone morphological transformations (erosion / dilation) in order to reduce the noise due to bubbles on the lower planes, (f) image on which the area A inside the black contours is measured and circles of radius $R = \sqrt{A / \pi}$ are shown, (g) original image on which the detected bubbles are represented as well as the size of the ``crop" in which the analysis was performed (red rectangle). The size distribution obtained is plotted in the center of the figure. }
 \label{fig:ImageAnalysis}
 \end{figure}

The image analysis of the data from ISS is difficult because of the following issues:

\begin{itemize}
\item The illumination is not homogeneous over the entire surface of the observed foam, as can be seen in the raw images of Figure \ref{fig:ImageAnalysis}(a). All the images show a fuzzy area. 
\item \DJD{Interior bubbles away from} the surface bubbles are seen in the images and should not be counted. 
\item The shape of the bubbles depends on the liquid fraction and evolves between $\phi$ = 15\% and 50\% from polyhedral to spherical shape. This requires having robust measurement algorithms working for all the observed shapes.
\item The bubbles are surrounded by a thick, inhomogeneous black outline. It is not obvious to decide where to measure the radius.  
Nevertheless, it is equivalent to choose either the middle of the black ring surrounding the bubbles or the inner contour of the bubbles. Our measurements show that the ratio between the radius of the inner contour and the radius in the middle of the black ring is constant and equal to 0.9. In the following, we use the radius measured from the middle of the black ring.
\item  The radius of bubbles seen in the images is smaller than the actual radius of the bubbles due to optical artifacts \cite{vanderNet2007}. This explains why the bubbles do not seem to touch, even at liquid fractions below the ``jamming" transition. 
\end{itemize}
For these reasons we tested different independent methods to extract the  bubble size distributions, and the resulting averages, from the images. In the next three  sections, we describe the different methods and we show in the fourth that they are indeed in very good agreement, thereby demonstrating the robustness of our measurements.

  \subsection{Manual image analysis\DJD{, using ellipses}}
  \label{sec:manual_analysis}
For every liquid fraction, raw images over the coarsening process have been analyzed ``by hand" using the software ImageJ. The contour of each bubble taken at the middle of the dark ring outlining the bubble is \Sylvie{manually} fitted by an ellipse (with small ellipticity, \textit{i.e.} between 1 and 1.15), and its area $A$ is measured. Then, the equivalent radius of  a circular bubble of the same area is calculated as $R = \sqrt{A / \pi}$. The systematic error that affects the measure of $R$ from the ISS images is estimated to be of the order of one pixel, \textit{i.e.} about $6~\mu$m. This image analysis allows the determination of the bubble radius distribution, since the earliest age just after the end of the foaming process until the end of the coarsening duration where there remains about 50 bubbles at the sample surface, for all investigated liquid fractions. From these effective radius distributions, several different moments are evaluated: the mean radius, the second and third moments $R_2=\left<R^2\right>^{1/2}$ and $R_3=\left<R^3\right>^{1/3}$ respectively, and the Sauter mean radius $R_{32}=\left<R^3\right>/\left<R^2\right>$.

 \subsection{Automatic measurement by standard image analysis\DJD{, using circles}}
 
 The image analysis has been performed using a Python script. The protocol for automated image analysis in Python is illustrated in Figure  \ref{fig:ImageAnalysis}: (a) raw image, (b) image whose brightness and contrast have been increased using the PIL library, (c) image whose contrast has again been improved and this time locally (using the \textit{equalize adapthist} function of \textit{Scikit-image}), (d) binarized image using a local thresholding method (adaptive threshold function of the CV2 library) (e) image having undergone morphological transformations (erosion / dilation) using the CV2 library in order to reduce the noise due to bubbles on the lower planes, (f) image on which the area $A$ inside the black contours is measured (using \textit{regionprops} from the \textit{Skimage} library) and circles of radius $R = \sqrt{A / \pi}$ are shown, (g) original image on which the detected bubbles are represented as well as the size of the ``crop" in which the analysis was performed (red rectangle). \Marina{Depending on the number of bubbles detected on an image, the threshold values are automatically adapted.} The example presented here corresponds to a foam with a liquid fraction of 40\% for which the distribution obtained is plotted in the center of the figure. The Sauter radius measured is (222 $\pm$ 58) $\mu$m with 290 bubbles detected. 
The results show that for each liquid fraction, the mean Sauter radius determined either ``by hand'' or by the Python algorithm described below coincide in the range comprised between $\approx 150~\mu$m and $\approx 1000~\mu$m.

\subsection{Automatic image analysis with machine learning}
The two previous subsections describe different methods used to analyze images of the surface bubbles. While both methods make accurate measurements the ``manual" elliptical fitting is time consuming and the ``automatic" circular fits work best for high liquid fraction foams; for both cases some bubbles are not well described by one single convex shape. To bypass these issues we turn to a machine learning algorithm for image segmentation called Bellybutton [https://pypi.org/project/bellybuttonseg/] \DJD{that is capable of handling a multiplicity of non-circular / non-ellipsoidal bubble shapes at any liquid fraction.} This algorithm uses a convolutional neural network trained on pixel-by-pixel inside vs outside classification of hand-segmented images; further details of the algorithm itself are beyond the scope of this paper. We do not use Bellybutton for any other analysis beyond the next subsection, where \DJD{we compare it with the other methods in order to judge the validity of all three approaches.}

\begin{figure*}[ht]
\centering
\includegraphics[width=6 in]{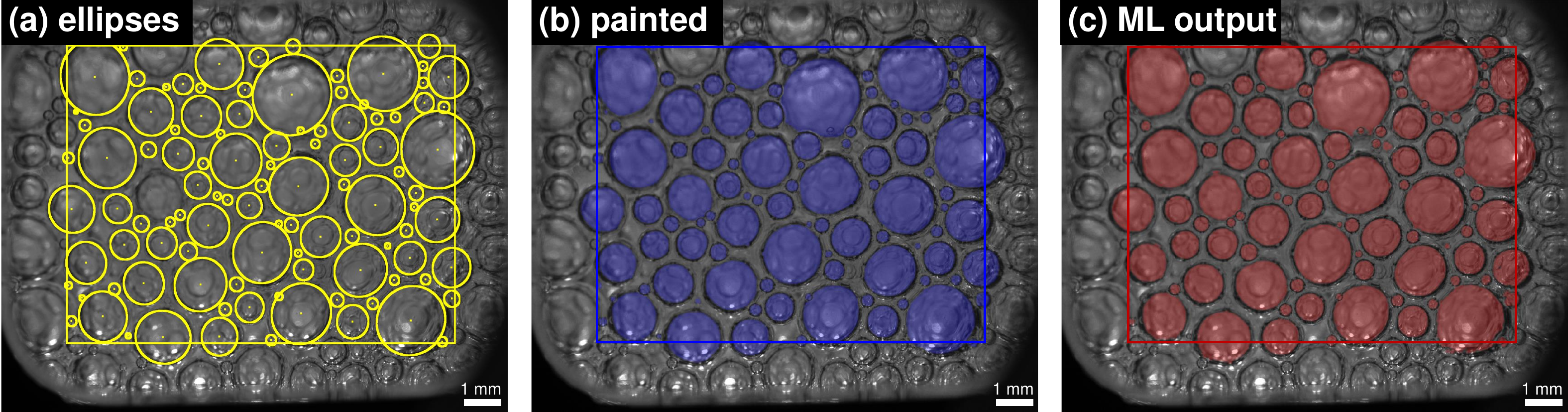} 
\caption{Three realizations of the surface bubbles on a single image of foam taken $t=29491$~s after initial preparation with 25\% liquid fraction. Parts (a) and (b) show the bubbles found by hand and the regions are either fits to ellipses or pixels chosen by hand to be in or out of a bubble, respectively. Part (c) shows regions identified by the machine learning algorithm.}
\label{fig:compare_realizations}
\end{figure*}

We train the algorithm on images where each pixel is determined by hand to be either inside or outside of a bubble. These pixels are chosen using a semi-transparent brush tool, and so we call bubbles identified via this method ``painted". \DJD{Note that such ``painting" represents yet a fourth image analysis method.} Fig.\ref{fig:compare_realizations}(b) shows an example of such a painted image. The machine learning algorithm works across various liquid fractions and it is independently trained on data from the 25\% and 40\% liquid fraction foams. We use 4 training images from the  25\% liquid fraction foam, $t=\{66,3604,25063,46542 \}$~s and 3 from 40\% liquid fraction, $t=\{18974,150251,657892\}$~s. Time is counted from preparation of the sample. \Anthony{Bellybutton will work regardless of liquid fraction but needs to be trained separately because of systematic changes in bubble features versus foam wetness.}

Once the algorithm is trained it is used to identify bubbles in images for the 25\% liquid fraction foams from, $t=\{68, 3941, 29491\}$~s and 40\% $t=\{42540,70789,361744\}$~s. \Anthony{Note that for the 25\% liquid fraction foams the images in the training set are from a different run of the experiment than the images from the prediction set.} Fig.~\ref{fig:compare_realizations}(c) shows an example output of the machine learning (ML) algorithm. We find excellent qualitative agreement with the underlying image. For the aforementioned sets of images we have both the machine learning output and an accompanying set of bubbles ``painted" by hand. For each image, we find the distribution of bubble areas generated via these two methods and compare them to the manual ellipse fits for the drier foams and the automatic circular fits for the wetter foams.

\Anthony{The dual purpose this subsection and the next is to demonstrate we can reconstruct the surface bubbles via machine learning, and to validate the other methods against ``painted" and machined-learned results. To show that machine learning works throughout the coarsening experiment we chose a subset of images spaced in time to capture representative features of the foams as the average bubble size increases. To predict bubble areas from a larger set of images or to improve accuracy we would need a larger training set albeit the ratio of 1 to 1 predicted images to training images would not be necessary. Needing a small number of training images for accurate predictions is useful because making the training set is time consuming. However an advantage of Bellybutton is the method of painting, training and predicting is the same for foams of any liquid fraction and can be generalized easily to other systems. Additionally it captures the details of the bubble shapes that are lost by using a single convex shape. }

\subsection{Comparison between the three methods}

\begin{figure}[ht]
\centering
\includegraphics[width=3.5in]{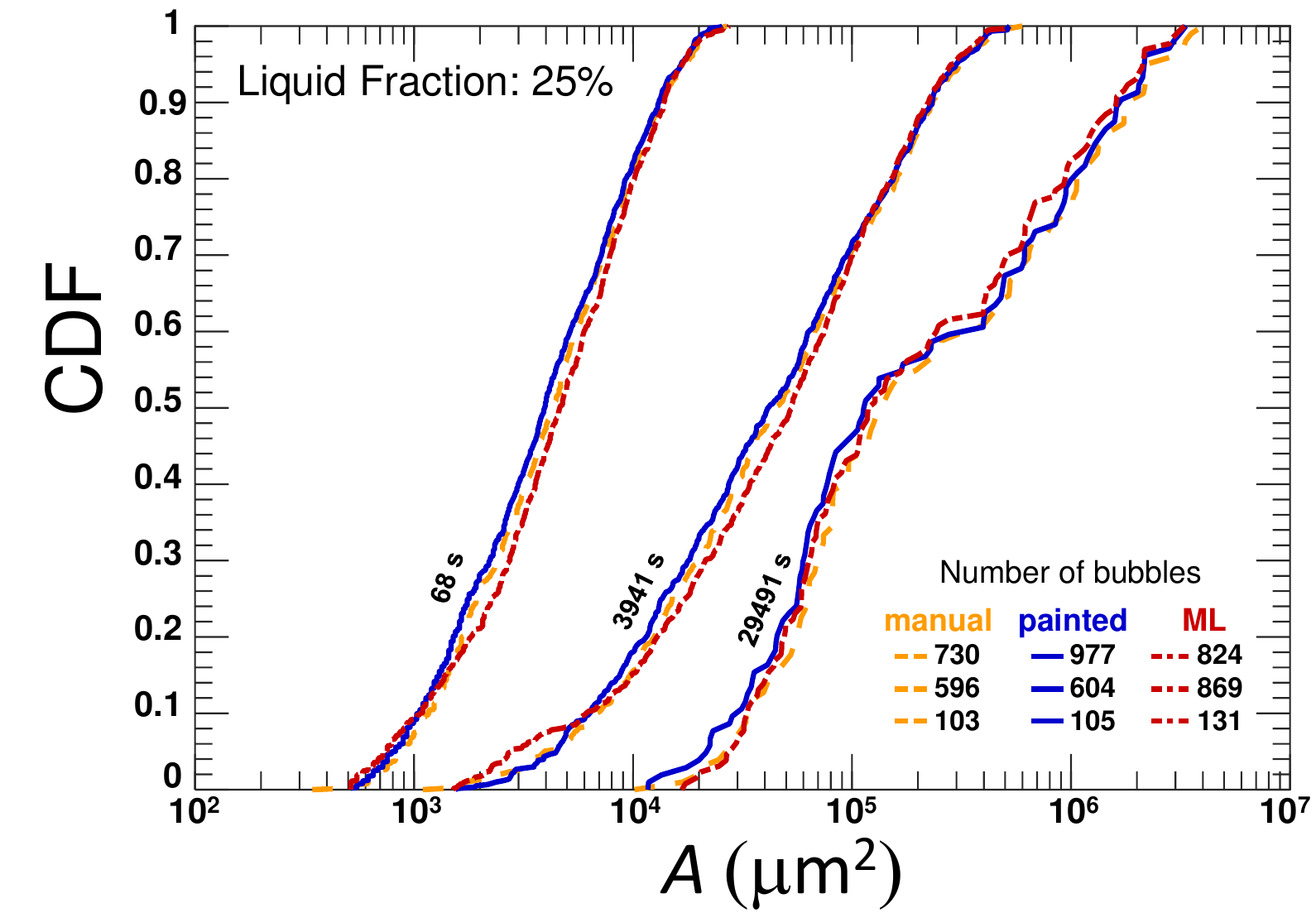} 
\caption{The cumulative distribution functions for coarsening surface bubbles at three different times with 25\% liquid fraction, as labeled. Areas are found by manual fits to ellipses, counting pixels in hand painted regions or by identifying regions with a machine learning algorithm. The curves for the ellipses, painted and machine learning (ML) data are shown using dashed, solid and dot-dashed lines, respectively.}
\label{fig:compare_CDF_25}
\end{figure}

To compare the different image analysis methods we use the cumulative distribution function (CDF) of the bubble areas. That is, for a given bubble area, the CDF value is the fraction of bubbles whose area is less than that value. Fig~\ref{fig:compare_CDF_25} shows the CDFs for the 25\% liquid fraction foams, using areas obtained from manual ellipse fits, painted bubbles, and machine learning. The first two methods identify areas by hand, and agree well with the output from Bellybutton. Deviations accrue for small bubbles in young foams using the machine learning output because of a population of bubbles just below the surface that are difficult to gather sufficient training data for, and that have shape characteristics similar to the actual surface bubbles. For older foams, where the surface bubbles are large and distinct from the background, all three methods have nearly identical distributions. 
 
\begin{figure}[ht]
\centering
\includegraphics[width=3.5in]{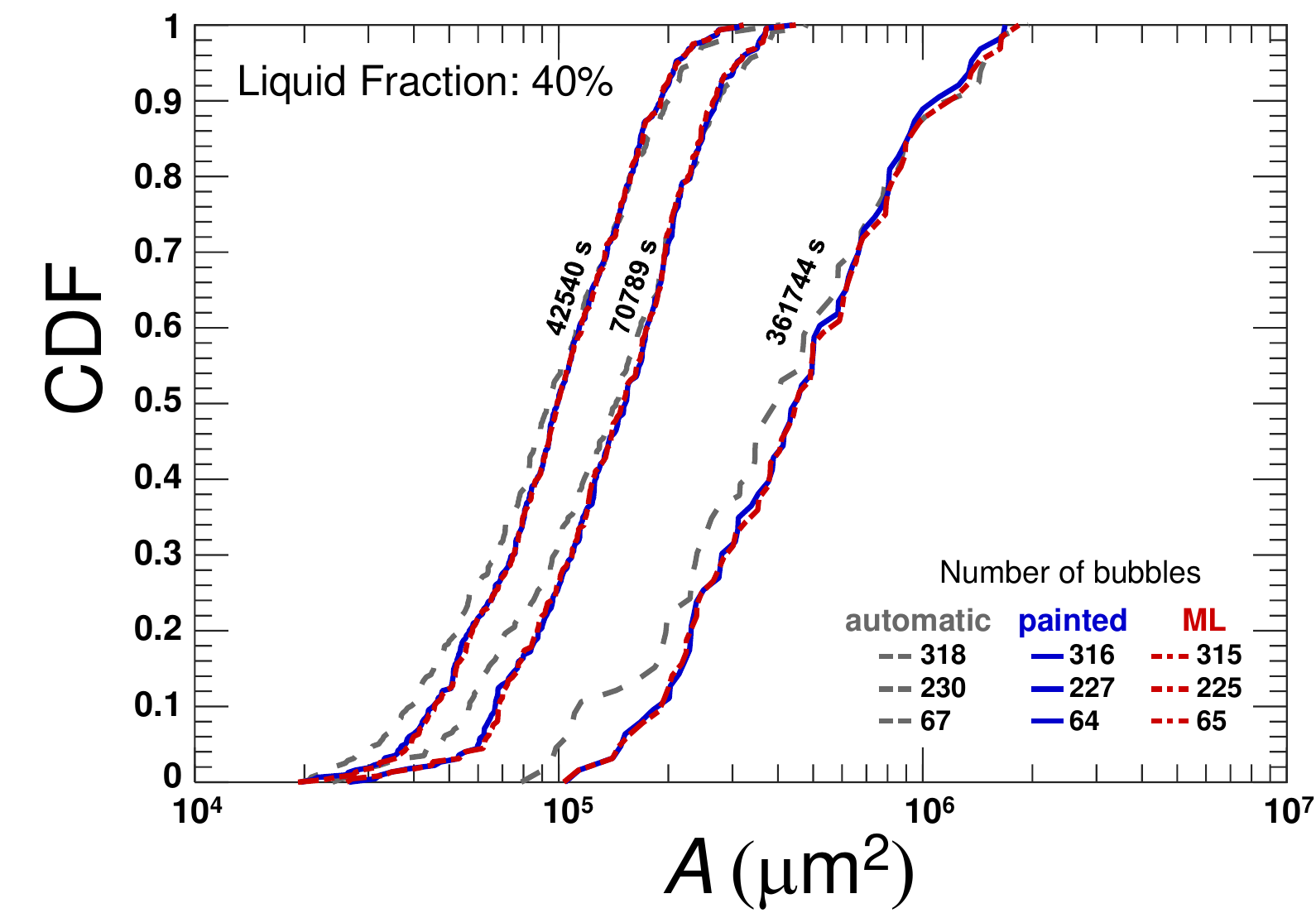} 
\caption{The cumulative distribution functions for coarsening surface bubbles at three different times with 40\% liquid fraction,as labeled. Areas are found by automatic fits to circles, counting pixels in hand painted regions or by identifying regions with a machine learning algorithm. The curves for the circles, painted and machine learning (ML) data are shown using dashed, solid and dot-dashed lines, respectively.}
\label{fig:compare_CDF_40}
\end{figure}

Fig.~\ref{fig:compare_CDF_40} shows the CDF for the 40\% liquid fraction foams at three different times. For the high liquid fraction foams the areas are collected from automatic circle fits, painted bubbles, and machine learning. We again see good agreement between the various methods. However there is a relative offset for the distributions found with the automated circle fits. This area difference is likely due to the constant offset found for the bubble radii. For these foams the agreement between the CDFs for bubbles painted by hand and bubbles found using the machine learning algorithm is nearly exact. \Anthony{Qualitative agreement between the CDFs found via the painted and Bellybutton methods is superior in Fig.~\ref{fig:compare_CDF_40} than in Fig.~\ref{fig:compare_CDF_25} but the distributions for the 40\% liquid fraction foams would lie to the right of even the latest time data in Fig.~\ref{fig:compare_CDF_25}. The latest time CDF for Fig.~\ref{fig:compare_CDF_25} is about as well resolved as all the CDFs in Fig.~\ref{fig:compare_CDF_40} and discrepancies for the former are likely due to whatever the algorithm learned in the training set.}

\DJD{Overall, the good comparison seen in Fig~\ref{fig:compare_CDF_25}-Fig~\ref{fig:compare_CDF_40} between the various methods demonstrates our ability to extract accurate bubble area distributions no matter what the liquid fraction.}

\section{Relation between bulk and surface bubble size distributions}

\label{sec:BulkSurface}
Due to successive reflections and refractions at the liquid-gas interfaces, foams strongly scatter light, preventing direct visualization of bubbles deep inside a bulk 3D sample. This effect becomes more and more pronounced as the liquid fraction increases. In our ISS experiments, where the investigated range of liquid fractions is comprised between 15\% and 50\%, the observations made with the overview camera are limited to the first layer of bubbles in contact with the top window (cf.~Fig.~\ref{fig:exmp}). This is also  the case on ground with  microscopy observation using similar illumination. Thus, the question arises whether the bubble size distribution measured at the surface of a 3D foam sample is representative \Sylvie{of} the bulk foam for the full range of investigated liquid fractions.
To address this issue, we have done a ground experiment to measure the bubble size distribution at the surface of a 3D foam sample together with the bubble size distribution of a monolayer of bubbles obtained by diluting the exact same foam sample with the foaming solution.

\begin{figure}
\begin{tabular}{cc}
 \begin{minipage}{0.3\textwidth} \includegraphics[width=0.9\textwidth]{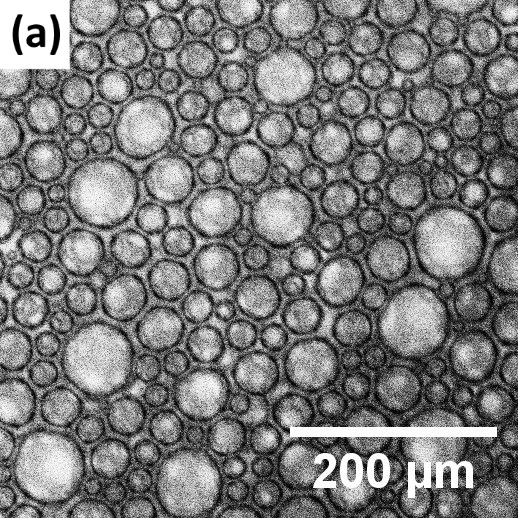} \\[0.1cm] \includegraphics[width=0.9\textwidth]{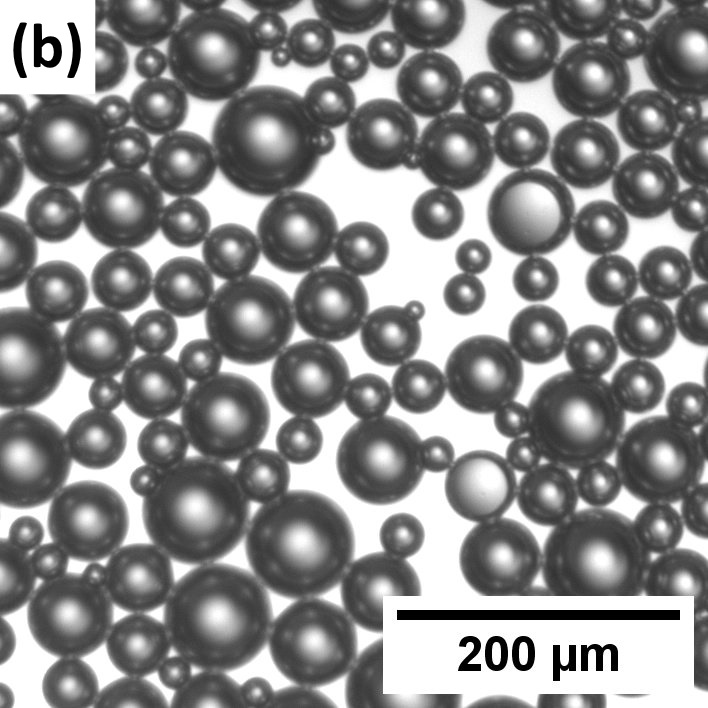} \end{minipage}
\begin{minipage}{0.7\textwidth} \includegraphics[width=0.99\textwidth]{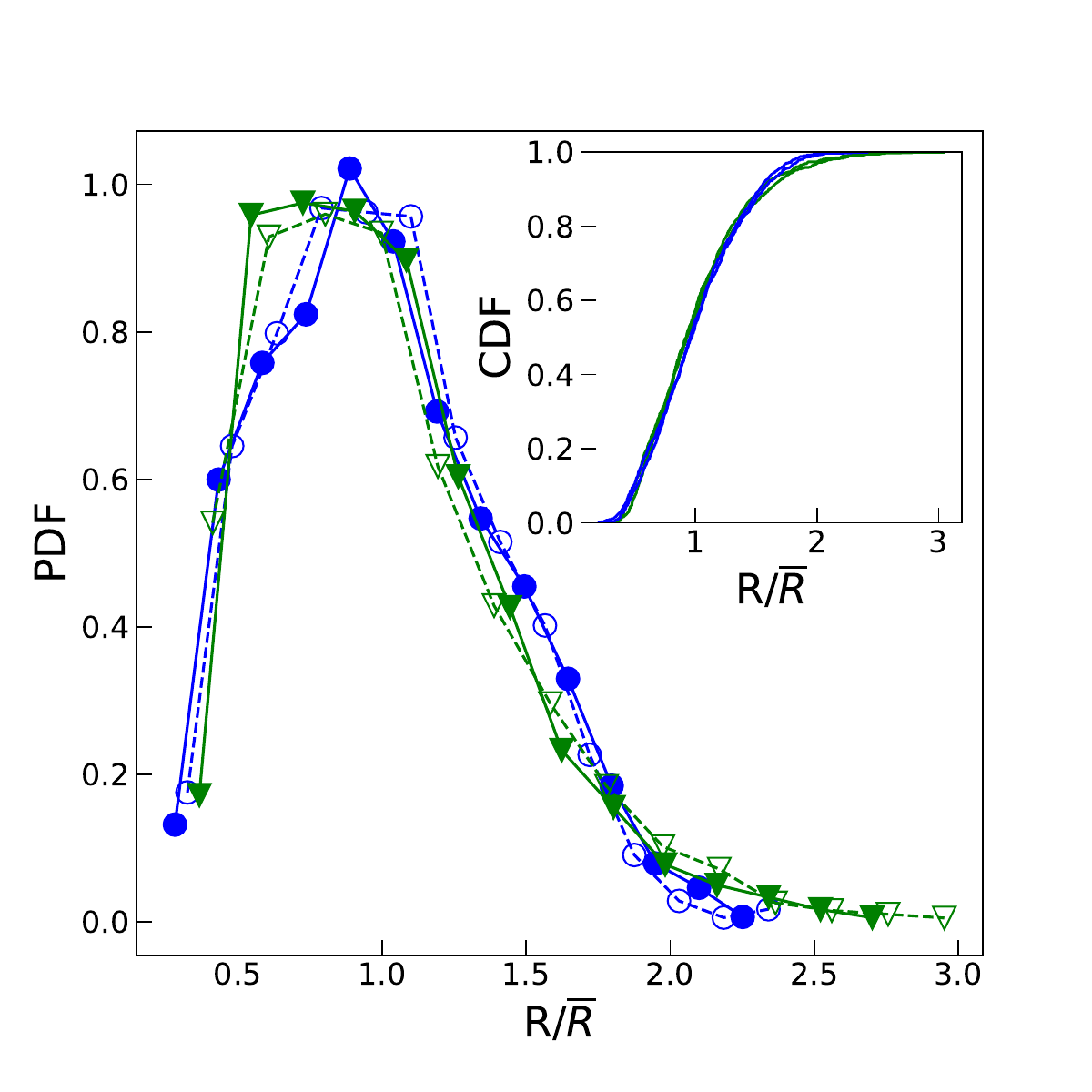} \end{minipage}& 
  \end{tabular}
        \caption{Images taken on ground of: a) the surface of a 3D foam with $\phi$ = 15\%, b) a  monolayer of bubbles of the same foam. c) Probability density functions of the radius distribution normalized by the average radius $\overline{R}$, for  \Nicolo{two} liquid fractions:  
   (\protect\tikz \protect\draw[blue,fill=blue] (0,0) circle (.8ex);, \protect\tikz \protect\draw[blue] (0,0) circle (.8ex);)  15$\%$ , (\protect\tikz \protect\draw[green, ultra thick, fill=green] (0.1,0) -- (0,0.2) -- (0.2,0.2)  -- cycle;, \protect\tikz \protect\draw[green, ultra thick] (0.1,0) -- (0,0.2) -- (0.2,0.2)  -- cycle;) 30$\%$ . 
   Filled markers refer to observations of bubbles in a monolayer, while empty markers correspond to bubbles at the surface of a 3D foam sample. Lines are guides for the eye. The inset shows the cumulative density functions of the same distributions.
  }
    \label{fig:surface_mono}
\end{figure}

The experimental set-up is described in section~\ref{sec:Experimentalground}, and two sample images are shown in Figure~\ref{fig:surface_mono}. We notice on the images taken at the foam surface (Fig.~\ref{fig:surface_mono}a) that the bubbles exhibit a dark contour similar to the one observed in the ISS images (Fig.~\ref{fig:ImageAnalysis}a). Indeed, since the slab thickness is more than 50~times larger than the mean bubble diameter, light is strongly scattered by the underlying bubbles in the foam so that the illumination conditions are similar for the foams in ground and in the ISS.
In contrast, as can be seen in Fig.~\ref{fig:surface_mono}b, bubbles in the monolayer appear on the microscopy images as dark disks delimited by a sharp contour. Since the average bubble size is in the range 10~$\mu$m $-$ 20~ $\mu$m, their Bond numbers are of the order $2.10^{-5}$ $-$ 1.2 $10^{-4}$, which means that the bubbles keep a spherical shape. Under these conditions, the analysis difficulties listed in Section \ref{sec:img_analysis} are prevented.

For each sample we take several pictures in different regions of the sample, to check for its homogeneity, and two pictures in the same place with 1~minute of time difference, to check the absence of evolution of the bubble size, either at the foam surface or in the bubbles monolayer. We conclude that on this short timescale, neither drainage nor coalescence nor coarsening modify the structures, so that the images of the monolayer are representative of the bubble population in the same foam sample from which they have been extracted. The images are analyzed with ImageJ by manually fitting ellipses on each bubble as described in section~\ref{sec:manual_analysis}. Between 1000 and 1200~bubbles per image are counted.

We did these experiments for  \Nicolo{two} liquid fractions: 15\% \Nicolo{and} 30\%. 
For each fraction, we measured the bubble radius distributions at the surface of the foam cell and for the bubble monolayer. We deduced the mean radius at the surface of the foam $\overline{R}$, the mean radius of the bubbles in the monolayer $\overline{R_m}$ which gives the ``real" mean radius of the bubbles that constitute the foam. \Sylvie{The mean radii $\overline{R}$ and $\overline{R_m}$ correspond to the first moment of the radius distribution.} In addition, the polydispersity index $p$ is evaluated either at the surface of the foam or for the monolayer. It is defined as $p =(\left<R^3\right>^{2/3} /\left<R^2\right>)-1$  \cite{Kraynik2004}. As reported in Table~\ref{tab:parameters}, we observe that the mean radius measured at the foam surface $\overline{R}$ is smaller than that of the monolayer $\overline{R_m}$. This means that such measurements at the surface underestimate the real bubble sizes. This is consistent with the fact that the dark contour that outlines the bubbles must be smaller than the actual bubble circumference. Indeed, due to a light refraction effect, the contact between two touching bubbles, as in the case for 15\% and  30\% liquid fractions, is not visible.  Based on a geometrical optics argument, the relationship between the contour radius $R$ and the real radius $R_m$ of an individual spherical bubble \Sylvie{in a monodisperse  ordered close compact packing} illuminated under diffuse transmitted conditions has been predicted to be~\cite{vanderNet2007}: $R = R_m  \cos ( \theta_c /2)$
where $\theta_c$ is the critical angle for total reflection between gas and liquid phases. For air-water interface with $\theta_c = 48.8^{\circ}$, this yields: $R \approx 0.91 R_m $. 

\Sylvie{The observed ratio $\overline{R} / \overline{R_m} $ is smaller than this value. The difference may be due to the effects of polydispersity, disorder and liquid fraction that prevent from making a direct comparison. }
Understanding the origin of the proportionality factor between the average radii $\overline{R}$ and  $\overline{R_m}$ as a function of the liquid fraction will be devoted to future experimental or ray-tracing numerical studies.
\par

\begin{table}[b]
    \centering
    \begin{tabular}{|c|c|c|c|c|}
    \toprule
        Liquid fraction $(\%)$
          & {$\overline{R_m}$ ($\mu m$)} & {$\overline{R}$ ($\mu m$)}  & $p_m$ & $p$  \\
        \midrule
        15 & $21.3\pm0.7$ & $16.2\pm0.9$ & $0.13$ & $0.12$ \\
        30 & $17.6\pm0.7$ & $13.0\pm0.9$ & $0.15$ & $0.17$\\
    \bottomrule
    \end{tabular}
    \caption{Mean apparent bubble radius $\overline{R}$ measured at the surface of the foam, mean bubble radius measured for the monolayer made of the same foam $\overline{R_m}$, and corresponding polydispersity indices $p$ and $p_m$ (defined in the text), for samples with different liquid fractions. The uncertainty on the mean radii is evaluated as a systematic error of 1 pixel size, i.e 0.7 $\mu$m or 0.9 $\mu$m for the bubble layer or the foam respectively.}
    \label{tab:parameters}
\end{table}

Figure~\ref{fig:surface_mono}c shows the probability density functions for the bubble radii, normalized by the mean value measured either at the surface of the foam sample or for the bubble monolayer. For a given liquid fraction, both bubble size distributions are similar. The cumulative distributions superpose remarkably well to each other. 
The results of the Kolmogorov-Smirnov statistical test \cite{KStest} on each couple of observations (monolayer and foam observed at surface) provide quantitative evidence for the similarity of the distributions, for 15$\%$ liquid fraction with a p-value~$=0.54$, and for 30$\%$ liquid fraction with a p-value~$=0.34$. 
\par
These results show that the  bubble size distributions, deduced from the outline contour of the bubbles at the surface of a 3D foam illuminated by diffuse transmitted light, are representative to a good approximation of the real bubble size distribution in the bulk of the sample, \Sylvie{in the illumination conditions of our experiment in the ISS}. Since the initial distributions of the foams produced in the ISS at the same liquid fractions are similar to these ones, we can infer that those distributions measured at the surface are representative of the bulk ones. 
Note that similar ground-based experiments cannot be performed on coarsened  \Nicolo{or wetter} foam samples, because gravity drainage becomes significant for these liquid fractions. This raises the question whether the coarsening induced evolution of the average bubble size is the same in the bulk and at the sample surface, an issue we address in section~\ref{sec:TDSanalysis}.

\section{A theory for the transmission of multiply scattered light through a sample cell of rectangular cuboid shape, with illumination and detection in small regions on opposite faces. \label{sec:transmissiontheory}}
 
Diffuse transmission spectroscopy is often used to analyze the average bubble size in  foams and granular materials  \cite{Hohler2014}. To implement this technique,  a slab shaped sample of thickness $L$ and a lateral extent much larger than $L$  is illuminated by an expanded beam of light and the ratio $T$ between the intensity   transmitted  through the sample and the incident one is measured. In the case of foams this involves many reflections and refractions at gas-liquid interfaces and can be represented as a  diffusive photon random walk with a transport mean free path   $\ell^*$. For samples where  $\ell^* \ll L$ and in the absence of light absorption, the fraction   of the incident intensity transmitted through a slab of thickness $L$ and infinite lateral extent follows the theoretical prediction \cite{Li1993, durian1991,Kaplan1994}:
\begin{equation}
 T \ =\ \dfrac{z_p +  z_e}{L/\ell^*  +   2 \; z_e}
 \label{Eq:Transmission2}
\end{equation}
\noindent
where $z_e$ is a parameter depending on the optical reflectivity of the walls of the slab, as explained in more detail below, and $z_p$  is the ratio between the average penetration depth \Reinhard{and $\ell^*$} at which the incoming light beam is scattered and converted into diffuse light. Experiments show that this ratio is  of order 1 \cite{Li1993}, in approximate agreement with theoretical models and simulations predicting $z_p =1$\cite{Lemieux1998}. In fact, it should be taken as $z_p=1$ unless there is significant amount of unscattered / ballistically transmitted light from the incident beam, in which case correction can be made that depend upon scattering anisotropy and boundary reflectvity \cite{Lemieux1998}.
The classic ``plane-in / plane-out" result of Eq.~\ref{Eq:Transmission2}, also equivalent to ``spot-in / plane-out" and ``plane-in / spot-out" where there is no discrimination based on lateral wandering of the photons, cannot be applied in our case because of many significant differences in illumination, light detection and lateral optical boundary conditions. 
Our sample has the shape of a rectangular cuboid, located in a region $-a/2 \le x \le a/2$,  $-b/2 \le y \le b/2$, $-L \le z \le 0$ where $x,y$ and $z$ are Cartesian coordinates and $a \sim b \sim L \gg \ell^{*}$. And we employ a ``spot-in / spot-out" illumination / detection geometry, where
the sample is illuminated by a narrow beam of light (diameter close to 1~mm) shining along the z~axis in the $-z$ direction and incident on the sample at the point $x=y=z=0$. The transmitted light is detected in a neighborhood of similar diameter on the opposite face, near the point  $x=y=0, \; z=-L$. We derive a theory for light transmission in a ``point-in point-out" configuration, as an approximation of our experimental spot-in / spot-out set-up. We expect this to be a good approximation, since the spot size is of the same order of magnitude as the limit of spatial resolution of the photon diffusion model, set by $\ell^*$. Moreover, the spot size is smaller than the lateral sample dimensions by an order of magnitude. Our theory is based on the diffusion approximation of radiative transfer theory \cite{Ishimaru1978}, assuming that light that has left the sample will not be backscattered towards it by other elements of the experimental setup. We will proceed in two steps: first we present a theory for light transmission through a slab of infinite lateral extent, in a "point-in / point-out" configuration. Then we introduce the lateral boundary conditions, to provide a realistic model that can be used for quantitative data analysis.\par
\Reinhard{\subsection{Classical analysis of diffuse light transmission through a slab of infinite lateral extent \label{sec:classical}}}
 The light  energy density $U$ inside the sample propagates as a dispersion of photons  that are  reflected and refracted by the gas-liquid interfaces. They thus do random walks, governed by a diffusion constant $D_L = v \ell^*/3$, where $v$ is the average speed of light propagation along the photon paths. As the collimated incident light beam  penetrates into the sample, \textit{i.e.} into the region $z<0$, more and more of its intensity is converted into scattered light which then propagates in random directions and is thus converted into ``diffuse light" \cite{Ishimaru1978}. For a random dispersion of point-like scatterers, the part of the incident light intensity which has not yet been scattered during its penetration  decreases exponentially with penetration distance $z$ according to the scattering length $\ell_s$ \cite{Ishimaru1978}. Note that taking a point or plane source at exactly $z_p=1$ is justified by \cite{Lemieux1998} for any degree of scattering anisotropy as long as there is essentially no unscattered = ballistically-transmitted light.    
 
 Fick's second law applied to photon diffusion with an exponentially decaying source yields the expression:
 \begin{equation}
 \label{eq:green1}
     D_L \; \Delta U = \frac{I s}{\ell} \;\delta(x) \; \delta(y) \; e^{z/\ell}
 \end{equation}
$I$ is the intensity of the incident beam, $s$ is its  cross-section area  and $\ell = z_p \ell^*$. The beam diameter is much smaller than $a,b$ and $L$. To simplify the mathematics, we model it as an infinitely thin beam,  described by the delta functions in Eq.~\ref{eq:green1}, represented  by the  symbol $\delta$. The symbol $\Delta $ represents the Laplace operator. To determine the photon density~$U$, we need to specify optical boundary conditions.
As a preliminary step, we  study the case of a slab  of infinite extent in the $x$ and $y$ directions, but bounded by transparent flat walls located at  $z=0$ and $z=-L$.
Photons near these walls can  easily leave the sample, so that here their concentration may be expected to be small. However, in reality the sample is bounded by reflecting transparent windows.  Some of the photons at $z\approx 0$ going in the direction of the $z$ axis are reflected back into the sample. The photon density at the boundary decreases as the boundary is approached from the inside of the sample, but it remains positive at $z=0$. A detailed analysis shows that in this case, the optical boundary condition must be characterized \Sylvie{with} an extrapolation of the function $U(z)$  to values of z outside the sample\cite{Ishimaru1978}. The distance between the point where this extrapolation reaches the value $U=0$ and the physical boundary is called "extrapolation length". The parameter $z_e$ is defined as this distance, normalized by $\ell^*$.  Mathematically, this boundary condition describing a partially reflecting interface is expressed by the following equations, relating $U$ and the gradient of $U$ in the outward direction, normal to the sample surface  \cite{Ishimaru1978, Durian1994}: 
\begin{align}
\label{eq:boundary1}
 z_e \ell^* \frac{\partial U({\bf r})}{\partial z}\bigg|_{z=0} +  U(x,y,0) &=0 \\
  - z_e \ell^* \frac{\partial U({\bf r})}{\partial z}\bigg|_{z=-L} +  U(x,y,-L) &=0.\label{eq:boundary2}
\end{align}
These are the correct boundary conditions;their approximate implementation in the case of complex sample geometries will be discussed in section \ref{sec:cuboid}.

\Reinhard{\subsection{Diffuse light transmission through a sample of infinite lateral extent, illuminated at a spot \label{sec:spotin}}}
\Reinhard{We will now construct a solution of Eq.~\ref{eq:green1} for a sample of infinite lateral extent and finite thickness, illuminated at a spot on one side. The case of a semi-infinite slab with illumination and detection at the same spot has been studied previously using the method of images}  \cite{Morin2002}. Here, we use the method of Green's functions, starting from the Poisson equation ~\cite{Jackson1975}
\begin{equation}
\Delta g({\bf r},{\bf r}_0) = \delta({\bf r}-{\bf r}_0) \label{eq:greena}
 \end{equation}
 It has the following solution, called Green's function:
 \begin{equation}
 \label{eq:green}
g({\bf r},{\bf r}_0) = -\frac{1}{4\pi |{\bf r}-{\bf r}_0|}
 \end{equation}
We can use these results to study light diffusion by linking the light energy density $U$ to $g$ as follows
\begin{equation}
    g({\bf r},{\bf r}_0)=\frac{D_L}{I s} U({\bf r}) \label{eq:UG}
\end{equation}
${\bf r}(x,y,z)$ represents the position of the observation point and ${\bf r}_0(x_0,y_0,z_0)$ the photon source position.
Eq.~\ref{eq:greena}  can thus be interpreted as a form of Fick's second equation in the presence of a  photon point source of intensity $I\,s$ located  in a uniform multiply scattering material of infinite extent. 
To adapt Eq.~\ref{eq:greena} to our boundary conditions Eq.~\ref{eq:boundary1} and Eq.~\ref{eq:boundary2}, we add to $g({\bf r},{\bf r}_0)$ a general solution of the Laplace equation $\Delta g({\bf r},{\bf r}_0) = 0$ which has a cylindrical symmetry \cite{Jackson1975}. This yields the following modified Green's function $G$, expressed in  cylindrical coordinates, which is the general solution of Eq.~\ref{eq:green1};
\begin{equation}
    G(\rho,z, z_0) =-\frac{1}{4\pi \sqrt{\rho^2 + (z-z_0)^2}} +  \int_0^{\infty}  (A(k) e^{k z}+ B(k) e^{- k z})  J_0(k\rho) dk
\end{equation}
$\rho$ is the radial cylindrical coordinate, $k$ is a wave number used in the Fourier representation of the Green's function  \cite{Jackson1975}.
 $J_0$ is the zero order Bessel function. The functions $A(k)$ and $B(k)$ must be determined so that the photon energy density $U$, proportional to $G$ according to Eq.~\ref{eq:UG},  fulfills the boundary conditions Eq.~(\ref{eq:boundary1}) and~(\ref{eq:boundary2}).
To achieve this,   we multiply these equations by $\rho J_0(k'\rho)$ and we integrate over $\rho$ from zero to infinity. Using the orthogonality of the Bessel functions~\cite{Jackson1975},
\begin{equation}
    \int_0^\infty\rho J_0(k\rho) J_0(k'\rho) d\rho = \frac{1}{k}\delta(k-k')
\end{equation}
we deduce the functions $A(k)$ and $B(k)$: 
\begin{align}
   A(k)&=\frac{e^{-k z_0} ( 1 - k \ell^* z_e)(-1+ k\ell^* z_e+e^{2k (L+z_0)}(1+k \ell^* z_e))}{4\pi(e^{2kL}(1+z_e k \ell^*)^2-(1-z_e k \ell^*)^2)   }\\
     B(k)&=\frac{e^{-k z_0} ( 1 - k \ell^* z_e)(1+ k\ell^* z_e+e^{2k z_0}(-1+k \ell^* z_e))}{4\pi(e^{2kL}(1+z_e k \ell^*)^2-(1-z_e k \ell^*)^2)   } 
\end{align}
To make further progress we recall the general expression of the flux of light {\bf F} is given by Fick's first law:
\begin{equation}
\label{eq:Fick1}
   {\bf F} = - D_L \; {\bf\nabla} U
\end{equation}
It determines the flux at any position in the sample. In previous work, the reflected flux has been analyzed as a function of the distance $\rho$, and it was found to be in good agreement with experimental results obtained for aqueous foams \cite{Hoballah1998}. Here we focus on the flux $F_T$ transmitted through the sample at  $z=-L$: 

\begin{equation}
    F_T = D_L \frac{\partial U}{\partial z}\bigg|_{z=-L}
\end{equation}
For a point source of unit intensity located at ${\bf r_0}$, $U$ is given by $G$ via Eq.\ref{eq:UG}. The expression of $U$ for the exponentially distributed source can be derived from this using the principle of superposition. We multiply $G(\rho,z,z_0)/D_L$ by the source intensity at the depth $z_0$, given by the right-hand side of  Eq.~(\ref{eq:green1}), and we integrate over $z_0$, from $-L$ to $0$.

This yields the  flux transmitted though a slab at $z=-L$  as a function of the distance $\rho$ from the z axis.  
\begin{equation}
F_T^{slab}(\rho)=-  \frac{I s}{\ell } \frac{\partial }{\partial z}\bigg|_{z=-L}\, \int_{-L}^0  e^{z_0/\ell}G(\rho,z,z_0)dz_0
\label{eq:slab}
\end{equation}

 The calculation yields the following expression for the  flux transmitted through a slab of infinite lateral extent, illuminated at a point. It is a decreasing function of the distance $\rho$ from the z axis.
\begin{equation}
\label{eq:pointtransmission}
  F_T^{slab}(\rho) =Is\int_0^\infty \frac{k e^{-\frac{L}{\ell }} \left(-2 k e^{L \left(\frac{1}{\
\ell }+k\right)} (\ell +\ell^* z_e)+(\ell  k+1) e^{2 k L} \
(k \ell^* z_e+1)-(\ell  k-1) (k \ell^* z_e-1)\right)}{2 \pi  \
\left(\ell ^2 k^2-1\right) \left((k \ell^* z_e-1)^2-e^{2 k L} \
(k \ell^* z_e+1)^2\right)}J_0(k\rho)dk
\end{equation}
A similar result has been derived previously\cite{Li1993}, under the more schematic assumption that photons are injected at a fixed penetration depth, instead of the exponential distribution used in Eq. \ref{eq:green1}.
 
\Reinhard{ \subsection{Diffuse light transmission through a cuboidal sample, using point-in point-out illumination and detection\label{sec:cuboid}}}

The results obtained so far do not take into account the {\it lateral} optical boundary conditions in our ISS experiment: The sample is not infinite in the $x$ and $y$ directions, but contained in a rectangular cuboid cell, located in a region $-a/2 \le x \le a/2$,  $-b/2 \le y \le b/2$, $-L \le z \le 0$.  At $ y=-b/2$, the bottom wall is made of an opaque plastic material which backscatters most of the incident light. Such a diffusely backscattering wall does not allow any flux penetration so that the optical boundary condition is  $z_e=\infty$.  The three other lateral walls are transparent, with an extrapolation parameter  $z_e $ that has the same value as on the faces perpendicular to the $z$ axis. However, implementing this condition analytically is difficult in the present case. Instead, we use for these three lateral boundaries the boundary condition $z_e=0$ which has previously been discussed in the literature~\cite{weiss1998}.Physically, it means that all photons that reach the boundary are assumed to leave the sample directly so that the light energy density $U$ is zero here.  This simplification enables an analytical solution, inspired by the method of images well known in electrostatics :   To implement  an optical boundary with $z_e=0$ at a given plane, we introduce virtual sources such that there is an anti symmetry between the sources on either sides of it. This means that to each source with intensity $I$ corresponds a virtual source of intensity $-I$ on the other side. By construction, we then have a vanishing photon density in the plane, as required. Of course, negative photon densities are not physical, this is a purely mathematical device, borrowed from electrostatics where similar arguments are often used concerning negative and positive charges\cite{Jackson1975}. A similar method is used to model the optical boundaries  with $z_e = \infty$ at a plane.  We introduce virtual image sources such that there is a mirror symmetry between the sources on either sides of the mirror plane. By construction, the photon density is then symmetric with respect to the plane and its gradient normal to it must everywhere be zero in the plane, as required.\par
\begin{figure}[t]
 \includegraphics[scale = 0.5]{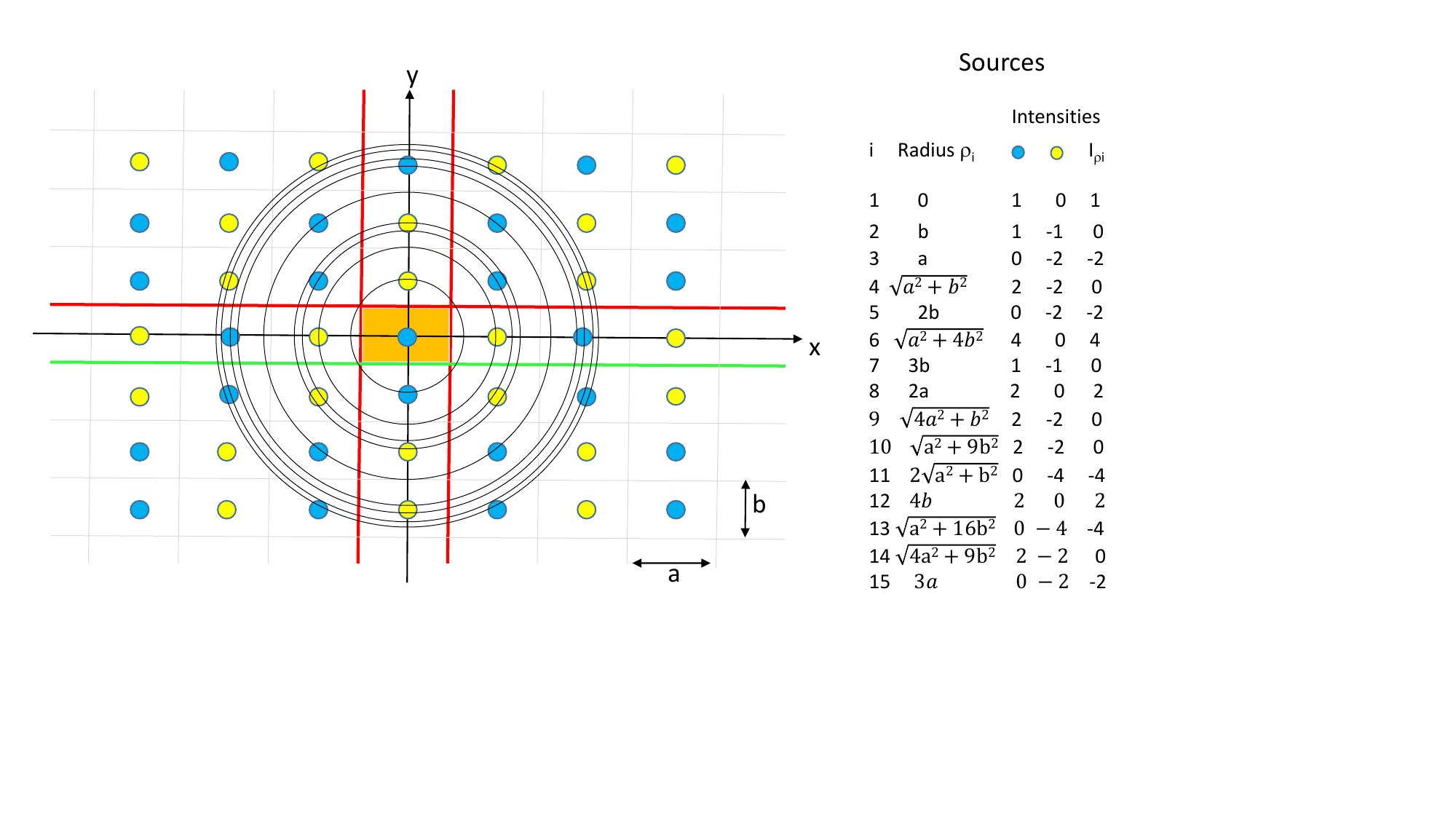}
 \caption{View of the illuminated face of the sample cell along  the direction of propagation of the incident laser beam i.e. to the $z$ axis in our calculations. Since the cell has a rectangular cuboid shape, its projection on the plane $z=0$ appears as  a rectangle (of dimensions $a$ x $b$) highlighted in orange at the center of the image.  The blue dot at  $x=0$, $y=0$ indicates the point like illumination by the incident beam of intensity $I$. To implement the optical boundary conditions at the sample faces perpendicular to the $x$ and $y$ axes, we consider a fictive sample which has the same thickness in the z direction as the physical one, but which is of infinite lateral extent. It is illuminated by virtual light sources of  intensity $+I$ (blue dots) and $-I$ (yellow dots) placed on a rectangular grid, at positions where $x$ and $y$ are respectively integer multiples of the lattice parameters $a$ and $b$. These sources are chosen such that the light intensity presents two planes of anti-symmetry perpendicular to the $x$ and $y$ axes, as discussed in the text. Their intersections with the plane $z=0$ are shown as a red line. There is also a plane of symmetry perpendicular to the $y$ axis whose intersection with the plane $z=0$, shown as green line. The virtual sources are placed at radial distances $\rho_i $ from the $z$ axis illustrated by circles and indicated  in the second column of the table on the right, as explained in the text.  For each radius $\rho_i$, the number of positive (blue) and negative (yellow) sources is given in the table, as well as their total intensity $I_{\rho i}$,  in units of the physical source intensity  $ I$.  \label{fig:sources} }
 \end{figure}

Figure~\ref{fig:sources} illustrates the set of sources needed to represent our specific ISS sample cell. For an arbitrary source location at $x=n a$ and $y = m b$, its distance from the origin is denoted $\rho = \sqrt{(na)^2+(mb)^2}$. 
The virtual source intensities are determined by the symmetries, they are given by the expression:
\begin{equation}
 (-1)^n \frac{\cos(m\pi/2 +\pi/4)}{|\cos(m\pi/2 +\pi/4)|}  s I.
\end{equation}
The set of sources is constructed by starting from the physical source at $x=y=0$ and by applying iteratively the required symmetries, as illustrated in figure~\ref{fig:sources}. The table next to the schema classifies groups of virtual sources, identified by the index $i$, depending on their distance $\rho_i$ from the physical source at the origin.  
Due to the symmetry of the setup, several sources often have the same distance to the origin. In this case, they share the same index $i$; the sum of all source intensities at a given $\rho_i$ that contribute to the transmitted flux $F_T^{cuboid}$ at $\rho=0, z=-L$, \Reinhard{ divided by $s I$}, is called $I_{\rho_i}$. We see that in several cases, the number of positive and negative sources at the same distance $\rho_i$ is equal so that their contribution is $I_{\rho_i}=0$.

\begin{figure}[h]
 \includegraphics[width=0.7 \linewidth]{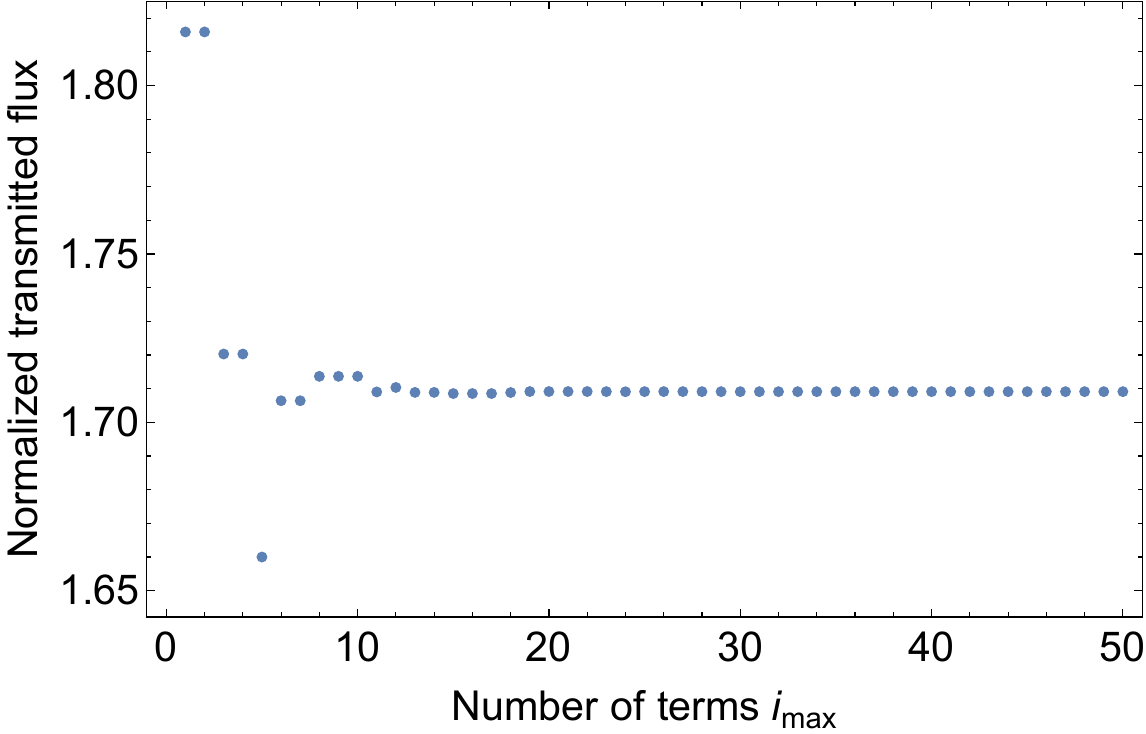}
 \caption{  $F_T^{cuboid}$ is the flux transmitted near a point at the sample surface opposite to the illuminated spot of a cuboid shaped sample. The plot shows the ratio of  $F_T^{cuboid}$ values, for transport mean free paths that can be encountered upon a coarsening experiment with a foam of liquid fraction 0.5:  $\ell^* =0.34 mm$, chosen as a reference value and $\ell^* =0.8 mm$. The vertical axis corresponds to  the ratio $F_T^{cuboid}/F_{T,o}^{cuboid}$ for these two $\ell^*$ values, predicted by Eq.\ref{eq:Ftb}.It is plotted versus the number of terms $i_{max}$ taken into account in Eq.\ref{eq:Ftb}. 
 The extrapolation length is $z_e=1.8$ for the investigated liquid fraction. The sample dimensions a, b, L considered in the calculation are those of our experimental configuration given in the text.      The plot shows that the convergence is almost complete for $i_{max}=15$. Similar convergence results are obtained for other liquid fractions and transport mean free path evolutions encountered in our coarsening experiments.} 
 \label{fig:convergence}
 \end{figure}

In figure~\ref{fig:sources}, the quantities  $I_{\rho_i}$ and $\rho_i$ are indicated up to $i= 15$.  
The intensity detected experimentally at $\rho = 0 $ is then predicted  as follows, using the principle of superposition:
\begin{equation}
F_T^{cuboid}=  \sum_{i=1}^{i_{max}} I_{\rho_i} \; F_T^{slab}(\rho_i) .  
\label{eq:Ftb}
\end{equation}
The number of terms in this sum is by construction infinite. To check the convergence of the sum and the number of terms required for an accuracy better than $1 \%$ we have calculated $F_T^{cuboid}$ as a function of the maximum value of   $i_{max}$. An example of such a result plotted in Fig.~\ref{fig:convergence} illustrates that the convergence is rapid; for all cases relevant in our experiments, we found that including additional terms beyond $i=15$ and up to 50  has an impact on the result far below $1 \%$.

  \begin{figure}[h!]
 \includegraphics[width=0.77 \linewidth]{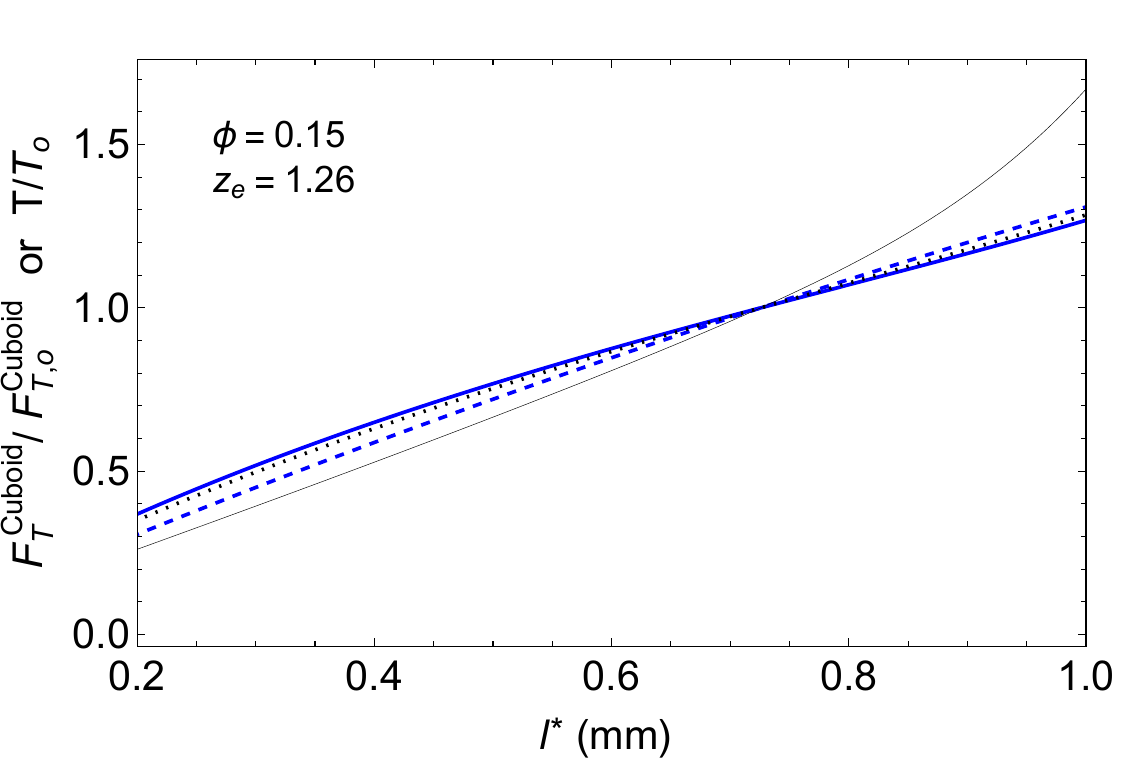}
 \includegraphics[width=0.74 \linewidth]{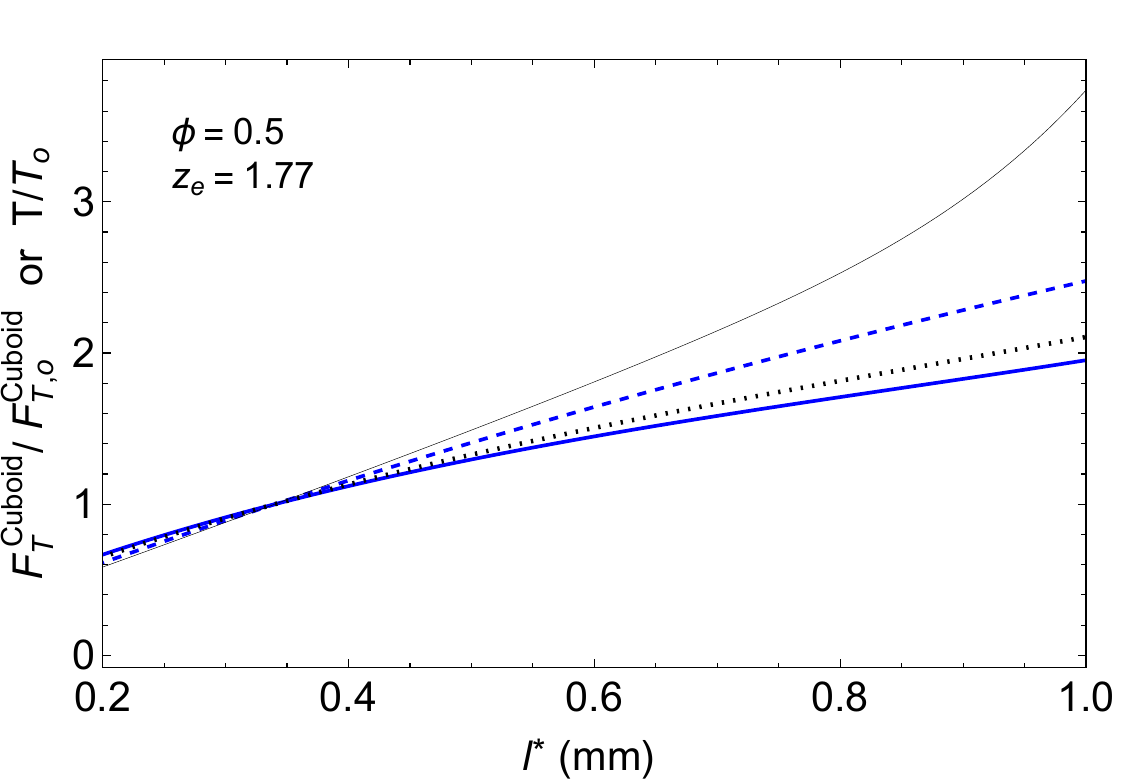}
 \caption{This plot illustrates the ratio of the transmitted flux  at a position opposite to the illuminated point on the cuboidal sample, $F_T^{Cuboid}$, for a range of transport mean free paths. It is normalized by its value $F_{T,o}^{Cuboid}$ for the reference transport mean free path $\ell^*_o$. It is chosen as in our experimental data analysis, $\ell^*_o$=~0.73~mm for $\phi=0.15$ and $\ell^*_o=$~0.34~mm for $\phi=0.5$. The two plots show predictions for these two different liquid fractions with corresponding extrapolation lengths $z_e$, as indicated. The full blue lines illustrate  Eq.~\ref{eq:Ftb}, using $i_{max}=15$ terms, thus taking into account the lateral boundary conditions of the cuboid. The dotted line illustrates the same quantity using only the first term in Eq.~\ref{eq:Ftb}. This means that a laterally infinite slab is modelled, without lateral boundaries. The gray thick line is a prediction for this latter case, assuming a boundary condition with extrapolation length $z_e=0$, illustrating the impact of the extrapolation length. The dashed line corresponds to the total transmission $T/T_o$ through a laterally infinite slab, normalized in the same way as $F_{T,o}^{Cuboid}$, and plotted versus $\ell^*$.}
 \label{fig:comparison1550}
 \end{figure}

 \Reinhard{
Fig. 9  provides an overview of the predictions obtained in this section. We consider the dependence on the transport mean free path of  the transmitted light flux at the center of the sample cell surface opposite to the one that is illuminated. This quantity, denoted $F_T^{Cuboid}$, is proportionnal to the photon count we measure experimentally in our setup.  Since we calibrate our experimental data by the transmission measured at a reference value of $\ell^*$, denoted $\ell^*_o$, we divide the predicted $F_T^{Cuboid}$, by its value predicted for $\ell^*=\ell^*_o$. This ratio is plotted on Fig. 9  as a function of $\ell^*$ for two different liquid fractions, 0.15 and 0.5, and the corresponding values of the extrapolation length $z_e$. The dotted lines on  Fig. 9 show the variation predicted for the point-in point-out configuration and a laterally infinite slab. This is an exact solution of the light diffusion equation where the boundary conditions 3 and 4 are fully implemented. As expected, the transmitted flux increases with  $\ell^*$.  
On Fig.  9 we also show the prediction of 17, where 15 terms are considered, this takes into account the lateral boundaries, assuming $z_e=0$ here. We recall that this $z_e$ value implies that any photon reaching the lateral boundary will be lost, while in reality some photons are reflected back at the interface. The calculation with $z_e=0$ provides an upper bound to the loss in transmission measured by  $F_T^{Cuboid}$ due to lateral boundaries. Fig. 9 shows that for our sample geometry the correction due to boundaries with respect to the unbounded point-in spont-out geometry is always less than 7 percent. It is much smaller than this in the dryer foam and for the smallest investigated transport mean free paths.}
\par \Reinhard{
To investigate  the error made by assuming $z_e=0$ rather than realistic, larger values, we have also plotted  on Fig.  9   $F_T^{Cuboid}/F_{T,o}^{Cuboid}$ without lateral boundaries, for the point-in point-out configuation, assuming $z_e=0$. For small transport mean free paths, the modified value of $z_e$ does not change the predicted transmission significantly, compared to our prediction for the same configuration with realistic $z_e$ values which are also shown, but a large increase appears for the biggest transport mean free paths.
By analogy with this result for laterally infinite samples, we  conclude that choosing $z_e=0$ for the lateral boundaries overestimates the photon loss and the reduction of $F_T^{Cuboid}/F_{T,o}^{Cuboid}$  it induces, seen on Fig. 9. An exact implementation of the lateral boundary conditions would therefore yield a prediction slightly above the full blue lines in 9, but well below the dotted ones. In view of other sources of small errors that we cannot control, we consider these corrections,  of the order of a percent, as negligible.}
\\
\Reinhard{We now discuss to what extent the transmitted flux in our experimental configuration is related to the classical theory of diffuse light transmission presented in section 5.1. It predicts the coefficient of tranmission $T$ (Eq.1) through the same kind of foam, but for illumination and detection all over the infinitely extended input and output surfaces. To make a connection with this configuration, we first discuss the case where only a  small spot is illuminated on the input side,and where we study the light intensity on the output side. It will reach a maximum at the point directly opposite to the illuminated spot and it will descrease with the lateral distance from this point. It can be shown that $T$, given by Eq.1, also describes the ratio between the integral over this heterogeneous transmitted flux and the illuminating light power.  However, we are not aware of any rigorous physical argument showing that this integrated intensity should scale with $\ell^*$ in the same way as  the flux in the center, given by $F_T^{Cuboid}$, for arbitrary boundary conditions. To investigate such a hypothetical relation empirically, we plot $T(\ell^*)/T(\ell^*_0)$ on  Fig. 9. The predicted increases of $T(\ell^*)/T(\ell^*_0)$ and $F_T^{Cuboid}/F_{T,o}^{Cuboid}$ are similar in the dryer foam, but there is a discrepancy up to $25 \%$ in the case of the wetter foam. Therefore, the ratio $T(\ell^*)/T(\ell^*_0)$ predicted by Eq.1  cannot be used to deduce the evolution of $\ell^*$ from our experimental data quantitatively.  }
\\

 \section{Analysis of Diffuse Transmission Spectroscopy data }
 \label{sec:TDSanalysis}
 To clarify whether surface and bulk bubbles follow the same average growth law, we measure the average bubble radius in the bulk of the foam, in parallel with the surface observations, for all the investigated liquid fractions. Using the ISS set-up, the diffuse intensity transmitted through the foam sample is measured during the course of coarsening (cf.~section~\ref{sec:ExperimentalISS}).  This allows the determination of the evolution of the photon transport mean free path $\ell^*$ (cf.~section~\ref{sec:transmissiontheory}), from which we deduce the average bulk radius along the coarsening, for any given liquid fraction.\par
 
 In the ISS set-up, the foam sample is contained in a cell of rectangular cuboid shape with thickness $L = 11.3$~mm, and lateral dimensions $a= 14.1$~mm and $b= 9.0$~mm, delimited by transparent walls on three sides and an opaque diffusing wall 
at the bottom due to the surface of the piston used for generating the foam.  Moreover, the sample is  illuminated at the center of one of the faces, and the multiply scattered light is collected via an optical fiber at the center of the opposite face. Therefore, the classic result (Eq.~\ref{Eq:Transmission2}) that relates the transmission coefficient to $\ell^*$ and $L$ for a slab of infinite lateral extent cannot be applied here. Instead, Eq.~\ref{eq:pointtransmission} and~\ref{eq:Ftb} 
provide the analog of the classic equation for this specific "point-in point-out" illumination configuration, cell shape and lateral reflectivity conditions.\par

The dependency of the extrapolation length $z_e$ with the foam liquid fraction is known from previous experiments~\cite{Vera2001}: $z_e$ increases from 0.88 for dry foams (air/glass/air interfaces) up to 1.77 for wet foams (water/glass/air interfaces). Since the refractive index of COC materials ($n=1.53$) that constitute the cell walls is very close to that of glass, \st{the same relationship can be used here} \Sylvie{the values of $z_e$ given in~\cite{Vera2001} are used here}. Previous experiments, using foams with different liquid fractions and bubble sizes have shown that~\cite{Vera2001}: $ \ell^* $ is proportional to the second moment of the bubble radius distribution $R_2$ observed at the sample surface, with a prefactor that decreases with increasing liquid fraction such as: $\ell^*~\approx~(3.0 + 0.28/ \phi) R_{2}$.

The ISS experimental set-up does not allow absolute measurements of the  transmitted intensity $F_{T}^{cuboid}$. Therefore, a calibration must be done for each experiment with a given sample, using a reference couple of a known transport mean free path and the corresponding transmitted intensity, respectively denoted $\ell^*_o$ and $F_{T,o}^{cuboid}$ . The reference $\ell^*_o$ is deduced from the average radius $R_2$ measured at the surface at a coarsening age $t_o$ when the foam has reached the scaling state regime, using the relationship mentioned above. The scaling state is characterized by the temporal invariance of the bubble size distributions measured at the surface (data to be published in a forthcoming paper). In this regime, the bubble size distribution is characterized by a single independent characteristic length scale, and any $n$-th moment $R_n$ of the radius distribution is related to the mean radius $\overline{R}$ by~\cite{Mullins1986, Hoballah1997}: $R_n \propto \overline{R}^{\;n} $. Since the factor of proportionality between $\ell^*$ and $R_2$ should be constant for a given bubble size distribution, the surface calibration should then not introduce any bias in the determination of $\ell^*$. 

For a given coarsening sample,  we do a numerical interpolation to predict  $\ell^*$ as a function of the measured normalized intensity $F_T^{cuboid}/ F_{T,o}^{cuboid}$, using Eq.~\ref{eq:pointtransmission} and~\ref{eq:Ftb}, assuming  $\ell= \ell^*$, as discussed above, and taking into account the first 15 terms of  the summation, from the indices $i=1$ to 15 as described in figure~\ref{fig:sources}. This allows a determination of $\ell^*$ along the coarsening. In practice, the measured transmission signal exhibits strong fluctuations towards the end of the coarsening experiment due to the presence of bubbles much bigger than the average that are encountered along the light paths. We therefore consider transmission measurements only when $L/\ell^* \gtrsim 10$. \Sylvie{Moreover, for consistency with the model proposed in section~\ref{sec:TDSanalysis}, we restrict our analysis to the size range where $b/\ell^* \gtrsim 10$ and $a/\ell^* \gtrsim 10$.}
Then we deduce the values of the mean Sauter radius $R_{32}$  assuming that $R_{32} \propto \ell^* $ which is consistent in the scaling state. 

\begin{figure}[h]
\centering
\includegraphics[width=18cm]{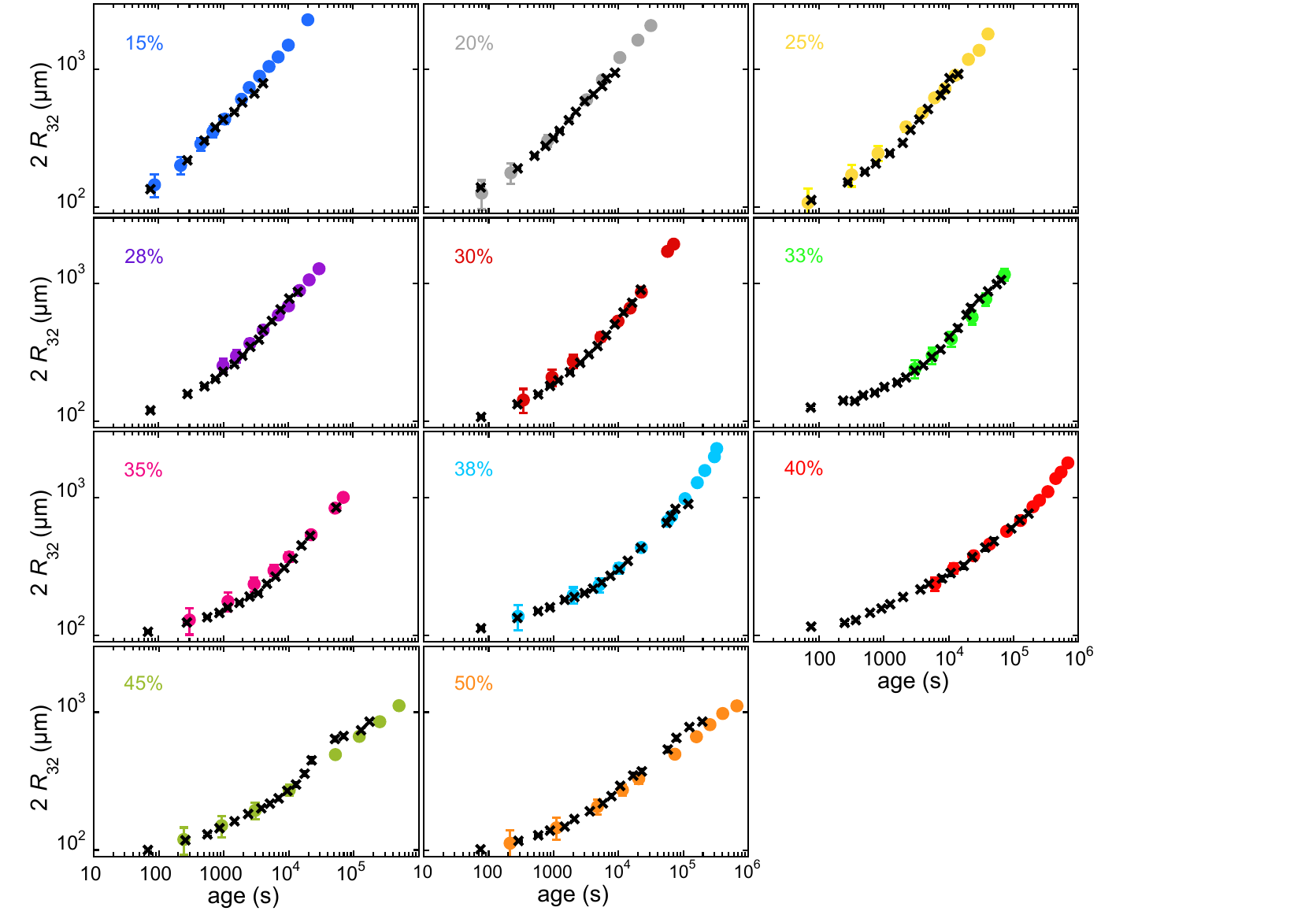} 
\caption{\Sylvie{ISS data: }Sauter mean diameter $R_{32}$ as a function of foam age, measured either in the bulk \Sylvie{of foam} using Diffuse Transmission Spectroscopy (crosses) or at the \Sylvie{foam} surface using videomicroscopy (disks). \Sylvie{The Sauter mean radii measured at the foam surface are deduced from the distributions of the radii of the contour of the bubbles, as explained in section~\ref{sec:manual_analysis}. The ages $t_o$ chosen for the calibration reference of the DTS measurements are: $t_o= 1050$ s for $\phi=15\%$, $t_o= 3230$ s for $\phi=20\%$, $t_o= 3940$ s for $\phi=25\%$, $t_o= 4110$ s for $\phi=28\%$, $t_o= 9860$ s for $\phi=30\%$, $t_o= 10960$ s for $\phi=33\%$, $t_o= 22500$ s for $\phi=35\%$, $t_o= 22520$ s for $\phi=38\%$, $t_o= 24300$ s for $\phi=40\%$, $t_o= 10150$ s for $\phi=45\%$, $t_o= 4730$ s for $\phi=50\%$.} The error bars over the surface data, shown unless of the size of the symbols, represent measurement uncertainties. Statistical uncertainty on DTS measurements are smaller than the size of the symbols. The liquid volume fraction is labelled inside each graph. }
\label{fig:Bulk_vs_Surface}
\end{figure}

Fig.~\ref{fig:Bulk_vs_Surface} shows the temporal evolution of the Sauter mean radius determined for all the investigated liquid fractions. The excellent superposition of these evolutions to those of $R_{32}$ measured at the surface constitutes a strong indication that the last ones are indeed representative of the bulk  bubbles'  evolution. For the two largest liquid fractions, we observe in fig.~\ref{fig:Bulk_vs_Surface} a small fluctuation of $R_{32}$ in bulk at the latest ages. Since the samples are polydisperse, bubbles larger than average may sit in a position close to the axis of incident beam propagation. Such big bubbles  will transmit more efficiently light, which yields a locally higher transmitted intensity, and hence corresponds to a larger  $\ell^*$ value, and in turn a larger local bubble size.  These fluctuations could also arise from local regions much less compact than the rest of the sample.  
Note that at the short ages, when the bubble size distribution has not yet reached its scaling state form, both geometrical coefficients $\ell^*/  R_{2}$ and $R_{32} / R_{2}$ may be slightly different from their asymptotic values, and exhibit a small dependency with polydispersity which may slightly modify the bulk radius values. However, this does not prevent the analysis of the average bubble growth laws in the scaling state. 

Light ray propagation through bubble packings raises further open theoretical questions, such as the relationship between $\ell^*$ and the different moments of the particle size distribution or the relation between the penetration depth  $z_p$  and the foam structure. These challenging questions are beyond the scope of the present paper. 

 \section{Conclusion}

Extracting the bubble size distribution from images of a foam surface  is difficult and full of pitfalls.   We have used three different  procedures: manual analysis, automatic analysis with a customized Python script and machine learning analysis. Remarkably, once the various pitfalls were identified as described in the paper, all the three procedures yielded identical results within error bars.
Additional ground experiments were performed to compare bubble size distributions at the surface and in the bulk of wet foams right after their production. The size distributions were found to be, to a good approximation, the same. 

To investigate whether the coarsening dynamics of the foam structure in the bulk is the same as at the sample surface, we have analyzed the transmission of diffuse light through the sample in microgravity. This technique, called
Diffuse Transmission Spectroscopy provides an average bubble radius in the bulk of the foam as a function of time. 

The cuboid shape and the ``point-in / point-out" illumination and light detection geometry of our setup are far from the standard case analyzed in the literature, where a slab shaped sample of infinite lateral extent is considered, with   illumination and detection of transmitted light all over the two opposite surfaces.

Using the theory of diffuse light transport we derive an analytic expression relating the transport mean free path to the measured transmitted intensity in the course of the coarsening experiments, fully taking into     account our experimental configuration. Using this, the temporal evolution of the bulk average radius  deduced from the transmitted intensity is \Dominique{the same to a factor probably close to 1 than that} of the average radius of the bubbles measured at the sample surface by image analyses.
It is to be noted that the predicted relation between transmitted intensity and transport mean free path can be generalized to other types of cuboid cell shapes or boundary reflectivities, as well as  other types of dispersions such as emulsions, colloidal particles or granular materials. 

\par

In conclusion, the tools provided by the ISS module are efficient and well adapted to foams in the context of a microgravity environment, but difficulties were encountered during the data analysis. They were solved as described in this paper and we \Dominique{were} thus able to investigate
the physics of bubble coarsening in wet foams, which will be in a forthcoming publication
 
  \section{Acknowledgements}
  
We acknowledge funding by ESA and CNES (via the projects “Hydrodynamics of Wet Foams”), as well as NASA via grant number 80NSSC21K0898. Marina Pasquet and Nicolo Galvani benefited from CNES PhD grants. The authors are grateful to the BUSOC team for their invaluable help during the ISS experiments. We also want to warmly thank Marco Braibanti from ESA and Olaf Schoele-Schulz from Airbus for their continuing support.

\bibliographystyle{unsrtnat}
\bibliography{references} 

\begin{thebibliography}{24}
\providecommand{\natexlab}[1]{#1}
\providecommand{\url}[1]{\texttt{#1}}
\expandafter\ifx\csname urlstyle\endcsname\relax
  \providecommand{\doi}[1]{doi: #1}\else
  \providecommand{\doi}{doi: \begingroup \urlstyle{rm}\Url}\fi

\bibitem[Cantat et~al.(2013)Cantat, Cohen-Addad, Elias, Graner, Höhler,
  Pitois, Rouyer, and Saint-Jalmes]{Cantat2013}
Isabelle Cantat, Sylvie Cohen-Addad, Florence Elias, François Graner, Reinhard
  Höhler, Olivier Pitois, Florence Rouyer, and Arnaud Saint-Jalmes.
\newblock \emph{Foams: Structure and Dynamics}.
\newblock Oxford University Press, Oxford, 2013.
\newblock ISBN 978-0-19-966289-0.
\newblock \doi{10.1093/acprof:oso/9780199662890.001.0001}.
\newblock URL \url{http://ukcatalogue.oup.com/product/9780199662890.do#}.

\bibitem[Gibson and Ashby(1997)]{Gibson1997}
L.~J. Gibson and M.~F. Ashby.
\newblock \emph{Cellular solids}.
\newblock Cambridge University Press, 2nd edition, 1997.

\bibitem[Mullins(1986)]{Mullins1986}
W.~W. Mullins.
\newblock The statistical self‐similarity hypothesis in grain growth and
  particle coarsening.
\newblock \emph{Journal of Applied Physics}, 59\penalty0 (4):\penalty0
  1341--1349, 1986.
\newblock \doi{10.1063/1.336528}.
\newblock URL \url{https://doi.org/10.1063/1.336528}.

\bibitem[Taylor(1998)]{Taylor1998}
P.~Taylor.
\newblock Ostwald ripening in emulsions.
\newblock \emph{Advances in Colloid and Interface Science}, 75\penalty0
  (2):\penalty0 107--163, 1998.
\newblock ISSN 0001-8686.
\newblock \doi{https://doi.org/10.1016/S0001-8686(98)00035-9}.
\newblock URL
  \url{https://www.sciencedirect.com/science/article/pii/S0001868698000359}.

\bibitem[Born et~al.(2021)Born, Braibanti, Cristofolini, Cohen-Addad, Durian,
  Egelhaaf, Escobedo-Sánchez, Höhler, Karapantsios, Langevin, Liggieri,
  Pasquet, Rio, Salonen, Schröter, Sperl, Sütterlin, and
  Zuccolotto-Bernez]{Born}
P.~Born, M.~Braibanti, L.~Cristofolini, S.~Cohen-Addad, D.~J. Durian, S.~U.
  Egelhaaf, M.~A. Escobedo-Sánchez, R.~Höhler, T.~D. Karapantsios,
  D.~Langevin, L.~Liggieri, M.~Pasquet, E.~Rio, A.~Salonen, M.~Schröter,
  M.~Sperl, R.~Sütterlin, and A.~B. Zuccolotto-Bernez.
\newblock Soft matter dynamics: A versatile microgravity platform to study
  dynamics in soft matter.
\newblock \emph{Review of Scientific Instruments}, 92\penalty0 (12):\penalty0
  124503, 2021.
\newblock \doi{10.1063/5.0062946}.

\bibitem[Wang and Neethling(2009)]{Wang2009}
Yingjie Wang and Stephen~J. Neethling.
\newblock The relationship between the surface and internal structure of dry
  foam.
\newblock \emph{Colloids and Surfaces A: Physicochemical and Engineering
  Aspects}, 339\penalty0 (1):\penalty0 73--81, 2009.
\newblock ISSN 0927-7757.
\newblock \doi{https://doi.org/10.1016/j.colsurfa.2009.01.021}.
\newblock URL
  \url{https://www.sciencedirect.com/science/article/pii/S0927775709000636}.

\bibitem[Cheng and Lemlich(1983)]{Cheng1983}
H.~C. Cheng and R.~Lemlich.
\newblock Errors in the measurement of bubble-size distribution in foam.
\newblock \emph{Industrial \& Engineering Chemistry Fundamentals}, 22\penalty0
  (1):\penalty0 105--109, 1983.
\newblock ISSN 0196-4313.
\newblock \doi{10.1021/i100009a018}.
\newblock Cheng, hc lemlich, r.

\bibitem[Durian et~al.(1991)Durian, Weitz, and Pine]{durian1991}
D.~J. Durian, D.~A. Weitz, and D.~J. Pine.
\newblock Multiple light-scattering probes of foam structure and dynamics.
\newblock \emph{Science}, 252\penalty0 (5006):\penalty0 686--688, 1991.
\newblock \doi{10.1126/science.252.5006.686}.
\newblock URL
  \url{https://www.science.org/doi/abs/10.1126/science.252.5006.686}.

\bibitem[Gaillard et~al.(2017)Gaillard, Roché, Honorez, Jumeau, Balan,
  Jedrzejczyk, and Drenckhan]{Gaillard2017}
T.~Gaillard, M.~Roché, C.~Honorez, M.~Jumeau, A.~Balan, C.~Jedrzejczyk, and
  W.~Drenckhan.
\newblock Controlled foam generation using cyclic diphasic flows through a
  constriction.
\newblock \emph{International Journal of Multiphase Flow}, 96:\penalty0
  173--187, 2017.
\newblock ISSN 03019322.
\newblock \doi{10.1016/j.ijmultiphaseflow.2017.02.009}.

\bibitem[Cohen-Addad et~al.(2004)Cohen-Addad, Hohler, and
  Khidas]{Cohen-Addad2004}
S.~Cohen-Addad, R.~Hohler, and Y.~Khidas.
\newblock Origin of the slow linear viscoelastic response of aqueous foams.
\newblock \emph{Physical Review Letters}, 93\penalty0 (2):\penalty0 028302,
  2004.
\newblock \doi{10.1103/PhysRevLett.93.028302}.

\bibitem[van Der~Net et~al.(2007)van Der~Net, Blondel, Saugey, and
  Drenckhan]{vanderNet2007}
A~van Der~Net, L~Blondel, A~Saugey, and W~Drenckhan.
\newblock Simulating and interpretating images of foams with computational
  ray-tracing techniques.
\newblock \emph{Colloids and Surfaces A: Physicochemical and Engineering
  Aspects}, 309\penalty0 (1-3):\penalty0 159--176, 2007.

\bibitem[Kraynik et~al.(2004)Kraynik, Reinelt, and van Swol]{Kraynik2004}
A.~M. Kraynik, D.~A. Reinelt, and F.~van Swol.
\newblock Structure of random foam.
\newblock \emph{Phys Rev Lett}, 93\penalty0 (20):\penalty0 208301, 2004.
\newblock ISSN 0031-9007 (Print) 0031-9007 (Linking).
\newblock \doi{10.1103/PhysRevLett.93.208301}.
\newblock URL \url{https://www.ncbi.nlm.nih.gov/pubmed/15600978}.

\bibitem[Höhler et~al.(2014)Höhler, Cohen-Addad, and Durian]{Hohler2014}
Reinhard Höhler, Sylvie Cohen-Addad, and Douglas~J. Durian.
\newblock Multiple light scattering as a probe of foams and emulsions.
\newblock \emph{Current Opinion in Colloid $\&$ Interface Science}, 19\penalty0
  (3):\penalty0 242--252, 2014.
\newblock \doi{http://dx.doi.org/10.1016/j.cocis.2014.04.005}.
\newblock URL
  \url{http://www.sciencedirect.com/science/article/pii/S1359029414000430}.

\bibitem[Li et~al.(1993)Li, Lisyansky, Cheung, Livdan, and Genack]{Li1993}
J.~H Li, A.~A Lisyansky, T.~D Cheung, D~Livdan, and A.~Z Genack.
\newblock Transmission and surface intensity profiles in random media.
\newblock \emph{Europhysics Letters ({EPL})}, 22\penalty0 (9):\penalty0
  675--680, jun 1993.
\newblock \doi{10.1209/0295-5075/22/9/007}.
\newblock URL \url{https://doi.org/10.1209/0295-5075/22/9/007}.

\bibitem[Kaplan et~al.(1994)Kaplan, Dinsmore, Yodh, and Pine]{Kaplan1994}
P.~D. Kaplan, A.~D. Dinsmore, A.~G. Yodh, and D.~J. Pine.
\newblock Diffuse-transmission spectroscopy: A structural probe of opaque
  colloidal mixtures.
\newblock \emph{Phys. Rev. E}, 50:\penalty0 4827--4835, Dec 1994.
\newblock \doi{10.1103/PhysRevE.50.4827}.
\newblock URL \url{https://link.aps.org/doi/10.1103/PhysRevE.50.4827}.

\bibitem[Lemieux et~al.(1998)Lemieux, Vera, and Durian]{Lemieux1998}
P.~A. Lemieux, M.~U. Vera, and D.~J. Durian.
\newblock Diffusing-light spectroscopies beyond the diffusion limit: The role
  of ballistic transport and anisotropic scattering.
\newblock \emph{Physical Review E}, 57\penalty0 (4):\penalty0 4498--4515, APR
  1998.
\newblock \doi{10.1103/PhysRevE.57.4498}.
\newblock URL \url{https://link.aps.org/doi/10.1103/PhysRevE.57.4498}.

\bibitem[{Ishimaru}(1978)]{Ishimaru1978}
A.~{Ishimaru}.
\newblock \emph{{Wave propagation and scattering in random media. Volume 1 -
  Single scattering and transport theory}}, volume~1.
\newblock Academic Press, 1978.
\newblock \doi{10.1016/B978-0-12-374701-3.X5001-7}.

\bibitem[Durian(1994)]{Durian1994}
D.~J. Durian.
\newblock Influence of boundary reflection and refraction on diffusive photon
  transport.
\newblock \emph{Physical Review E}, 50\penalty0 (2):\penalty0 857--866, 1994.
\newblock URL \url{http://link.aps.org/abstract/PRE/v50/p857}.

\bibitem[Morin et~al.(2002)Morin, Borrega, Cloitre, and Durian]{Morin2002}
F.~Morin, R.~Borrega, M.~Cloitre, and D.~J. Durian.
\newblock Static and dynamic properties of highly turbid media determined by
  spatially resolved diffusive-wave spectroscopy.
\newblock \emph{Applied Optics}, 41\penalty0 (34):\penalty0 7294--7299, DEC 1
  2002.
\newblock \doi{10.1364/AO.41.007294}.
\newblock URL \url{https://link.aps.org/doi/10.1364/AO.41.007294}.

\bibitem[Jackson(1975)]{Jackson1975}
John~David Jackson.
\newblock \emph{{Classical electrodynamics; 2nd ed.}}
\newblock Wiley, New York, NY, 1975.
\newblock URL \url{https://cds.cern.ch/record/100964}.

\bibitem[Hoballah(1998)]{Hoballah1998}
H.~Hoballah.
\newblock \emph{Disproportionnement, Structure et Rheologie d'une Mousse
  Aqueuse}.
\newblock PhD thesis, Université de Marne la Vallée, 1998.

\bibitem[Weiss et~al.(1998)Weiss, Porrà, and Masoliver]{weiss1998}
George~H. Weiss, Josep~M. Porrà, and Jaume Masoliver.
\newblock The continuous-time random walk description of photon motion in an
  isotropic medium.
\newblock \emph{Optics Communications}, 146\penalty0 (1):\penalty0 268--276,
  1998.
\newblock ISSN 0030-4018.
\newblock \doi{https://doi.org/10.1016/S0030-4018(97)00475-6}.
\newblock URL
  \url{https://www.sciencedirect.com/science/article/pii/S0030401897004756}.

\bibitem[Vera et~al.(2001)Vera, Saint-Jalmes, and Durian]{Vera2001}
Moin~U. Vera, Arnaud Saint-Jalmes, and Douglas~J. Durian.
\newblock Scattering optics of foam.
\newblock \emph{Appl. Opt.}, 40\penalty0 (24):\penalty0 4210--4214, Aug 2001.
\newblock \doi{10.1364/AO.40.004210}.

\bibitem[Hoballah et~al.(1997)Hoballah, Höhler, and Cohen-Addad]{Hoballah1997}
H.~Hoballah, R.~Höhler, and S.~Cohen-Addad.
\newblock Time evolution of the elastic properties of aqueous foam.
\newblock \emph{J. Phys. II France}, 7:\penalty0 1215--1224, 1997.

\end{thebibliography}

\end{document}